\documentclass[a4paper,11pt]{article}
\usepackage{epsfig}
\usepackage{axodraw}
\newcommand{\be}{\begin{equation}}
\newcommand{\en}{\end{equation}}
\newcommand{\bea}{\begin{eqnarray}}
\newcommand{\ena}{\end{eqnarray}}
\newcommand{\lbl}[1]{\label{eq:#1}}
\newcommand{\rf}[1]{(\ref{eq:#1})}
\newcommand{\dslash}{%
\mathrel{\setbox0=\hbox{$\partial$}\copy0\kern-0.8\wd0\hbox{\slash}}}
\newcommand{\Aslash}{%
\mathrel{\setbox0=\hbox{$F$}\copy0\kern-0.8\wd0\hbox{\slash}}}
\newcommand{\pslash}{%
\mathrel{\setbox0=\hbox{$p$}\copy0\kern-0.8\wd0\hbox{\slash}}}
\newcommand{\kslash}{%
\mathrel{\setbox0=\hbox{$k$}\copy0\kern-0.8\wd0\hbox{\slash}}}
\newcommand{\qslash}{%
\mathrel{\setbox0=\hbox{$q$}\copy0\kern-0.8\wd0\hbox{\slash}}}
\newcommand{\qqslash}{%
\mathrel{\setbox0=\hbox{$q'$}\copy0\kern-1.0\wd0\hbox{\slash}\kern0.5\wd0}}

\newcommand{\gapprox}{%
\mathrel{%
\setbox0=\hbox{$>$}\raise0.6ex\copy0\kern-\wd0\lower0.65ex\hbox{$\sim$}}}

\def\Ldvmu{{\log{\mu^2\over M_V^2}}}
\def\Ldvmuz{{\log{\mu_0^2\over M_V^2}}}
\def\LdvZ{{\log{M^2_Z\over M_V^2}}}
\def\Lz{\log(z)}
\def\mld{ {M_l^2} }
\def\ml {{M_l} }
\def\mg {{M_\gamma}}
\def\mgd{{M_\gamma^2}}
\def\mud{{\mu^2}}

\def\iddk{ {-id^4 k\over (2\pi)^4\,}}
\def\iddkd{ {-id^d k\over (2\pi)^d\,}}
\def\div{ {\rm div}}
\def\divexp{ {\log{\Lambda^2\over\mu_0^2}}}
\def\divchi{ {\rm div}_\chi }
\def\facpi{{1\over 16\pi^2\,}}
\def\mufacpi{{1\over 16\pi^2\,}}
\def\dilog{{\rm dilog}}
\def\LLeps{ \log {M_\gamma\over M_l\,}}
\def\LLdeps{ \log^2 {M_\gamma\over M_l\,}}
\def\gffL{{-{4G_F V_{ud}\over\sqrt2}}}
\def\gff{{-{  G_F V_{ud}\over\sqrt2}}}
\def\Qbar{ {\overline{Q}}}
\def\psibd{ {\overline{d}_L}}
\def\psiu { {u_L}}
\def\psibl{ {\overline{l}_L} }
\def\psil{ {l_L}}
\def\psinu{ {\nu_L}}
\def\psibnu{{\overline{\nu}_L}}
\def\ubarl{ {\bar u_l(p)}}
\def\nuL { {{1-\gamma_5\over2}v_\nu(q)}}
\def\mvd{{M_V^2}}
\def\mad{{M_A^2}}
\newcommand{\trace}[1]{\langle #1 \rangle}
\def\ql{\mathbf{q}_{L}}

\def\qr{\mathbf{q}_{R}}
\def\qrb{\mathbf{q}_{R^b}}

\def\qw{ \mathbf{q}_{W}}
\def\Qw{ {Q_W}}
\def\qwc{ \mathbf{q}_{W^c}}
\def\qv{ \mathbf{q}_{V}}
\def\qvb{ \mathbf{q}_{V^b}}
\def\qa{ \mathbf{q}_{A}}
\def\qab{ \mathbf{q}_{A^b}}
\def\qwdag{ {\mathbf{q}_W^\dagger}}
\def\qql{{{\cal Q}_L}}
\def\qqr{{{\cal Q}_R}}
\def\qqw{{{\cal Q}_W}}
\def\qqlmu{{{\cal Q}^\mu_L}}
\def\qqrmu{{{\cal Q}^\mu_R}}

\begin{document}
\begin{titlepage}
\begin{flushright}
IPNO/DR-05-03\\
LPT-ORSAY/05-29\\
\today
\end{flushright}
\begin{center}
{\Large\bf Radiative corrections}\\ 
{\Large\bf in weak semi-leptonic processes at low energy:}\\
{\Large\bf a two-step matching determination}
\footnote{Work supported in part by the EU RTN contract 
HPRN-CT-2002-00311 (EURIDICE).}\\[1.5cm]

{\large S. Descotes-Genon$^a$ and B. Moussallam$^b$} 

{\sl$^a$\ Laboratoire de Physique Th\'eorique\footnote{
LPT is an unit\'e mixte de recherche du CNRS et de l'Universit\'e Paris-Sud 11
(UMR 8627).} }\\
{\sl Universit\'e Paris-Sud 11, F-91406 Orsay, France}

{\sl$^b$\ Institut de Physique Nucl\'eaire\footnote{
IPN is an unit\'e mixte de recherche du CNRS et de l'Universit\'e Paris-Sud 11
(UMR 8608).} }\\ 
{\sl Universit\'e Paris-Sud 11, F-91406 Orsay, France}
\end{center}
\vfill

\begin{abstract}
%
We focus on the chiral Lagrangian couplings describing radiative corrections 
to weak semi-leptonic decays and relate them to the decay 
amplitude of a lepton, computed by Braaten and Li at one loop in 
the Standard Model.
For this purpose, we follow a two-step procedure. A first
matching, from the Standard Model to Fermi theory, yields a relevant
set of counterterms. The latter are related to chiral couplings thanks to
a second matching, from Fermi theory to the chiral
Lagrangian, which is performed using the spurion method.
We show that the chiral couplings of physical relevance obey
integral representations in a closed form, expressed
in terms of QCD chiral correlators and vertex
functions. We deduce exact relations among the couplings, as well 
as numerical estimates which go beyond 
the usual $\log(M_Z/M_\rho)$ approximation.
\end{abstract}
\vfill

\end{titlepage}
\section{Introduction}
An accurate evaluation of radiative corrections in $K_{l3}$ decays 
is crucial for a precise determination of $V_{us}$. In this context,
it is necessary to control whether experimental data on
$K^+_{l3}$ and $K^0_{l3}$ data are consistent~\cite{cirignp04}. 
Several new experiments have studied $K$ decays. Results on the
$K^+_{l3}$ mode were released very recently by the E865~\cite{E865}
and ISTRA~\cite{ISTRA} collaborations and results on the $K^0_{l3}$
mode were presented by the NA48~\cite{NA48}, KTev~\cite{KTev} and
KLOE~\cite{KLOE} collaborations. 
This has stimulated renewed interest in the theoretical determination of 
radiative corrections in such processes~\cite{cirig01,newrad}. 

This subject has a long history~\cite{kinoshita}. Within  the 
Standard Model, a conspicuous feature of radiative corrections to 
semi-leptonic decays is their enhancement by a large logarithm
$\log(M_Z/\mu)$ with $\mu\simeq 1$ GeV, which was pointed out 
by Sirlin~\cite{sirlin78,sirlin82}. 
In this paper, we will focus on the remaining
(unenhanced) corrections and will discuss a method for 
determining them.
The proper theoretical framework to discuss semi-leptonic decays of
kaons (as well as those of $\pi$'s or $\eta$'s) is 
the chiral effective Lagrangian
formalism~\cite{weinberg79,gl84,gl85} (see the book~\cite{dghbook}
for a review of applications).  
The discussion of radiative corrections
requires extensions of the original setting which were performed
successively by Urech~\cite{urech95} and then by 
Knecht \emph{et al.}~\cite{knrt}. 
At this stage, the effective Lagrangian includes not only the pseudo-Goldstone
bosons, but also the photon and the light leptons as dynamical degrees
of freedom. In other words, this Lagrangian describes the whole Standard
Model at low energies. High-energy dynamics has been integrated out into
local (contact) terms, parameterised by a set of low-energy constants (LEC's).

In this paper, we will consider the set of LEC's $X_i$ introduced
in ref.~\cite{knrt} to deal with virtual leptons and 
discuss their physical interpretation. In particular,
we will show that they satisfy simple integral representations in terms of
QCD Green functions in the chiral limit. These results extend those obtained
in the case of the Urech LEC's $K_i$ for virtual photons~\cite{bm97}, which
were themselves generalisations of the well-known sum rule by 
Das~\emph{et al.}~\cite{dgmly}. These integral representations 
provide practical means of estimating the LEC's $X_i$ numerically,
once the chiral Green functions are approximated by
simple, large-$N_c$ motivated, models. 
But our analysis goes beyond these numerical results, since we will derive
some non-trivial relations among the LEC's $X_i$ and with
the electromagnetic coupling $K_{12}$.
This will allow us to clarify completely a related issue, 
the dependence of $K_{12}$ on short-distance 
renormalisation conditions, observed in~\cite{bm97} and 
further discussed in~\cite{rusetsky}.
We start from a result of Braaten and Li (denoted BL 
in the following)~\cite{bl}, who
computed the amplitude for a lepton\footnote{The authors of ref.~\cite{bl} 
were chiefly interested in the case of the $\tau$ lepton. 
However, their result is general and it will be applied to the 
light leptons $e$ and $\mu$ here.}
decaying into a massless quark, 
a massless antiquark, and a neutrino at one loop in the Standard Model.
This computation completed earlier results obtained by 
Sirlin~\cite{sirlin82}.

We will follow a two-step matching procedure which can be sketched as:
\begin{center}
\begin{minipage}{12cm}
\begin{picture}(427,25)(0,0)
\put(10,0){\framebox(80,15){Standard Model}}
\put(90,7.5){\vector(1,0){30}}
\put(120,0){\framebox(70,15){Fermi theory}}
\put(190,7.5){\vector(1,0){30}}
\put(220,0){\framebox(100,15){Effective Lagrangian}}
\end{picture}
\end{minipage}
\end{center}
This two-step procedure will allow us to determine the 
implications of BL's calculation for the effective Lagrangian.
It turns out to be particularly convenient to introduce
Fermi theory as an intermediate stage,
in order to integrate out the high-energy dynamics of the Standard Model
in a transparent way. In addition, at this intermediate stage,
we can rely on a Pauli-Villars regularisation (applied to the photon
propagator) to tame divergences. This regularisation scheme offers the
attractive feature of remaining in four dimensions, and thus avoids
the well-known difficulties of dimensional regularisation when 
defining $\gamma_5$. This will prove particularly useful 
when we deal with chiral QCD correlators.

The plan of the paper is as follows. We begin by reconsidering the
one-loop calculation of radiative corrections to the semi-leptonic
decay of a light lepton in Fermi theory. The ultraviolet divergences are 
removed through a set of counterterms. Matching the one-loop amplitude
in the Standard Model and in Fermi theory yields constraints on the
values of the latter. Then, we re-express the counterterms in Fermi
theory to introduce spurion sources instead of the electric and weak
charge matrices. Using this new form, we perform the second
matching step and identify counterterms in Fermi theory and LEC's in the
chiral Lagrangian. This identification involves also chiral two- 
and three-point Green functions. Finally, integral relations are derived, 
which are exploited to obtain analytical relations among chiral LEC's
and numerical estimates based on large-$N_c$ models 
for the relevant chiral correlators.

\section{One-loop matching of Fermi theory and the Standard Model
revisited}
\subsection{Tree-level amplitude}

Following ref.~\cite{bl}, we consider the amplitude $T(p,q,p',q')$ 
for the semi-leptonic weak decay of a lepton into massless quark,
antiquark and neutrino
\be\lbl{amplit}
l(p)\to \bar u(q) + d(q')+\nu (p')\ .
\en
The usual kinematical variables are introduced
\be
s=(p-q)^2,\quad t=(p-p')^2,\quad u=(p-q')^2,\quad s+t+u=\mld\ .
\en

\begin{figure}[t]
\begin{center}
\begin{tabular}{ccc}
\begin{picture}(175,175)(25,0)
\ArrowLine(25,100)(100,100)
\ArrowLine(100,100)(175,150)
\SetWidth{2}
\ArrowLine(125,50)(200,50)
\ArrowLine(200,20)(125,50)
\SetWidth{1}
\DashArrowLine(100,100)(125,50){5}
\Text(50,110)[]{$l(p)$}
\Text(155,120)[]{$\nu(p')$}
\Text(95,75)[]{$W^-$}
\Text(175,60)[]{$d(q')$}
\Text(160,20)[]{$\bar{u}(q)$}
\end{picture}
& \qquad\qquad &
\begin{picture}(175,175)(25,0)
\ArrowLine(25,100)(100,100)
\ArrowLine(100,100)(175,150)
\SetWidth{2}
\ArrowLine(100,100)(200,80)
\ArrowLine(200,40)(100,100)
\SetWidth{1}
\GBoxc(100,100)(10,10){0.5}
\Text(50,110)[]{$l(p)$}
\Text(155,120)[]{$\nu(p')$}
\Text(175,95)[]{$d(q')$}
\Text(160,50)[]{$\bar{u}(q)$}
\end{picture}
\end{tabular}

\caption{Tree-level diagram for the semi-leptonic decay of a lepton
into a massless quark, antiquark and neutrino in the Standard Model (left)
and in Fermi theory (right).}
\label{fig:treelevel}
\end{center}
\end{figure}
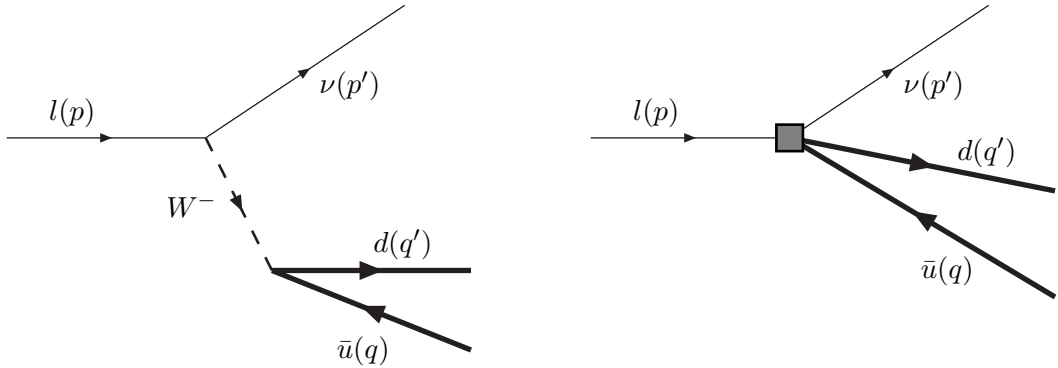

BL have computed the one-loop amplitude $T(p,q,p',q')$ in the Standard Model,
and we intend to perform the same work in Fermi theory in the presence of 
electromagnetic interactions (see fig.~\ref{fig:treelevel}).
The relevant part of the interaction Lagrangian is
\be\lbl{fermi}
{\cal L}_{Fermi}= \gffL \left\{
\psibl \gamma^\lambda\psinu  \times
\psibd  \gamma_\lambda\psiu   +h.c. 
\right\}\ ,
\en
At leading order, we consider the diagram in fig.~\ref{fig:treelevel},
which gives the following results for the amplitude
\be\lbl{T0}
T_0=\gff \bar u_\nu(p')\gamma_\lambda (1-\gamma^5) u_l(p)\,
\bar u_d(q')\gamma_\lambda (1-\gamma^5) v_u(q)\ ,
\en
and for the decay rate
\be\lbl{Gamma0}
\Gamma_0= {G_F^2 M_l^5\over192\pi^3} N_c V_{ud}^2\ .
\en
At this order, Fermi theory and the Standard Model yield identical results. 

\subsection{One-loop electromagnetic corrections in Fermi theory}

Let us turn to the one-loop corrections to this result. 
In the Standard Model, the decay rate 
$\Gamma$ receives contributions from exchanges of virtual
photons, weak gauge and Higgs bosons~\cite{sirlin78,bl}. 
Infrared divergences occur, 
but they are cancelled once we add the decay rate for real-photon 
emission $l\to \bar u+ d+\nu+\gamma$. In Fermi theory, 
the one-loop corrections which involve two weak vertices are negligibly
small and we only have to consider diagrams which involve 
the exchange of a photon between two charged fermion lines.
This contribution contains infrared divergences, which will be cancelled
by the decay rate for real-photon emission. At this order, the expression of
the latter is identical to that in the Standard Model.
In addition, starting at two loops (i.e. at order $O(\alpha\,\alpha_s)$ )
there appears QCD corrections to the decay amplitude.
One can use Fermi theory whenever the momentum transferred by the virtual
$W$ boson is much smaller than its mass. One must also require that
this momentum is sufficiently large as compared to 1 GeV such that perturbative
QCD makes sense. For the moment, let us  ignore these corrections.
In sec.~\ref{sec:secx6} we will discuss how to take them into account in an 
approximate way.

\begin{figure}[t]
\begin{center}
\begin{tabular}{ccc}
\begin{picture}(125,150)(50,25)
\ArrowLine(50,100)(100,100)
\ArrowLine(100,100)(175,150)
\Photon(160,52)(160,84){4}{3}
\SetWidth{2}
\ArrowLine(100,100)(175,80)
\ArrowLine(175,40)(100,100)
\SetWidth{1}
\GBoxc(100,100)(10,10){0.5}
\end{picture}
& 
\begin{picture}(125,150)(50,25)
\ArrowLine(50,100)(100,100)
\ArrowLine(100,100)(175,150)
\PhotonArc(110,100)(45,-20,180){4}{15}
\SetWidth{2}
\ArrowLine(100,100)(175,80)
\ArrowLine(175,40)(100,100)
\SetWidth{1}
\GBoxc(100,100)(10,10){0.5}
\end{picture}
& 
\begin{picture}(125,150)(50,25)
\ArrowLine(50,100)(100,100)
\ArrowLine(100,100)(175,150)
\PhotonArc(100,100)(30,180,320){4}{8}
\SetWidth{2}
\ArrowLine(100,100)(175,80)
\ArrowLine(175,40)(100,100)
\SetWidth{1}
\GBoxc(100,100)(10,10){0.5}
\end{picture}
\\
$Q_2 Q_3$ & $Q_0 Q_2$ & $Q_0 Q_3$
\end{tabular}

\caption{
One-loop electromagnetic corrections to the semi-leptonic decay of
a lepton in Fermi Theory. Diagrams corresponding to wave-function
renormalisation and proportional to $Q_i^2$ ($i=0,2,3$) are not
shown.}
\label{fig:oneloop}
\end{center}
\end{figure}
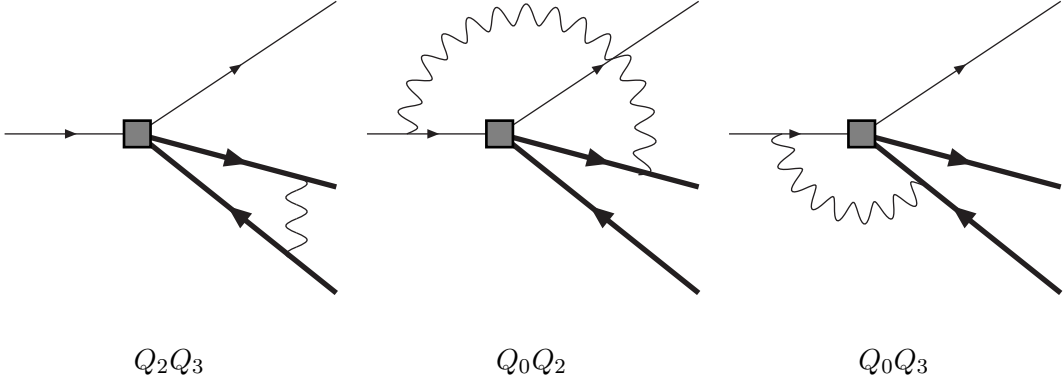

Therefore, we focus on $O(e^2)$ corrections to eq.~\rf{T0}
in Fermi theory caused by the exchange of a virtual photon.
The Lagrangian which encodes the interactions of the photon field
with the charged leptons and quarks is given by
\bea\lbl{gamma}
{\cal L}_{\gamma}=
&& -{1\over4}F^{\mu\nu}F_{\mu\nu} -{1\over2\xi} (\partial^\mu
F_\nu)^2 +{1\over2}\mgd F^\mu F_\mu +\nonumber \\
&& \bar{l} (i\dslash-eQ_0\Aslash-\ml) l 
  +\bar\nu_L    (i\dslash) \nu_L
  +\sum_{q=u,d}\bar{q}  (i\dslash-eQ_q\Aslash ) q \ . 
\ena
A small photon mass $M_\gamma$ is introduced in order to control infrared
divergences. In addition, we use the Pauli-Villars regularisation method to
treat ultraviolet divergences.
From the point of view of the chiral expansion the terms
in eq.~\rf{gamma} have chiral order $p^2$ provided that counting
rules are adopted
\be\lbl{count0}
F_\mu\sim O(p^0)\qquad 
l,\ \nu_L,\ q\ \sim O(p^{1\over2})\qquad
e,\ M_l,\ M_\gamma\sim O(p)\ .
\en
In this paper, we will restrict ourselves to the Feynman gauge $\xi=1$. 
Following BL's convention, we denote the charges of the lepton, the
quark and the antiquark $Q_0$, $Q_2$ and $Q_3$ respectively. The 
physical values of these charges are
\be
Q_0=-1,\ Q_2=-{1\over3},\ Q_3=-{2\over3}\ .
\en
Now, we determine the various contributions due to a virtual photon exchange,
labeled in terms of these charges and shown in fig.~\ref{fig:oneloop}. 

\subsubsection{Contributions $Q_0^2$ and $Q_2^2+Q_3^2$}\label{sec:self}

These contributions are given by the wave-function renormalisation.
Including contributions up to one loop, the lepton propagator
has the following form
\be\lbl{props}
G^l_F(p) ={i\over\pslash -\ml +\Sigma_l(p) }
\en
with
\bea
\Sigma_l(p)= -Q_0^2 e^2
\int\iddk {\gamma^\mu (\pslash+\kslash +\ml)\gamma_\mu
\over (k^2-\mgd)_\Lambda ((k+p)^2-\mld)}\ .
\ena
The denominator $(k^2-\mgd)_\Lambda$ stems from the photon propagator,
regularised \`a la Pauli-Villars
\be\lbl{pvprop}
{1\over(k^2-\mgd)_\Lambda}= 
{-\Lambda^2\over(k^2-\mgd)(k^2-\Lambda^2)} \ .
\en
The wave function renormalisation of the lepton requires to expand
the lepton propagator around the mass-shell $p^2=\mld$
\be\lbl{proplep}
G^l_F(p) \simeq {i\over (1+K^l_F)(\pslash -M_l-\delta M_l )} \ .
\en
A standard calculation gives
\bea
&& K^l_F= {-Q_0^2 e^2 \over16\pi^2}\left( -\div 
+2\log{M_l\over\mu_0}
-4\log{ \mg\over\ml}
-{9\over2} \right)\nonumber\\
&& \delta M_l={-Q_0^2 e^2 \over16\pi^2} M_l\left( 3\,\div 
-6\log{M_l\over\mu_0}
+{3\over2}\right)\ ,
\ena
with the (regularised) divergent piece
\be
\div= \log{\Lambda^2\over\mu_0^2} \ .
\en
and $\mu_0$ denotes the renormalisation scale in Fermi Theory.
Applying the LSZ reduction formula (e.g.~\cite{iz}) yields the correction 
of order $e^2 Q_0^2$ induced by the one-loop lepton propagator of the form~\rf{proplep}
\be
T_{00} = T_0 \left(-{1\over 2} K_l^F\right) =
T_0 \times {Q_0^2 \alpha\over 8\pi }
\left( -\div 
+2\log{M_l\over\mu_0}
-4\log{ \mg\over\ml}
-{9\over2}\right)\ .
\en

Quark propagators are treated on the same footing apart from the 
fact that these fermions are assumed to be massless. In this case, one finds
the wave-function renormalisation factor to be
\be
K^q_F= {-Q_q^2 e^2 \over16\pi^2}\left( -\div 
+2\log{\mg\over\mu_0}  \right)
\en
and the corresponding contribution to the decay amplitude reads
\be
T_{qq}= T_0\times (Q_2^2+Q_3^2) {\alpha \over 8\pi}
\left(   -\div 
+2\log{\mg\over\mu_0}  \right) \ .
\en

These yield the following corrections to the decay rate
\begin{equation}
\Gamma_{ii}=\Gamma_0 \frac{\alpha}{2\pi}
 \left[
  Q_0^2 \left( -\frac{1}{2}\div 
   +\log{M_l\over\mu_0}
   -2\log{ \mg\over\ml}
   -{9\over4}\right)
  +(Q_2^2+Q_3^2) \left(-\frac{1}{2}\div +\log{\mg\over\mu_0}  \right)
 \right]
\end{equation}

\subsubsection{Contribution $Q_2 Q_3$}

Here one considers the graph with one photon line 
joining the anti-quark $u$ to the
quark $d$ (left-hand diagram in fig.~\ref{fig:oneloop}). 
The amplitude has the form
\be
T_{23}= \gff \bar u_\nu(p')\gamma_\lambda (1-\gamma^5) u_l(p)
H^\lambda_{23}(q,q')\ ,
\en
where $H^\lambda_{23}$ is given by
\be
H^\lambda_{23}(q,q')=
Q_2 Q_3\, e^2\,
\bar u_d (q') \gamma^\mu\gamma^\sigma\gamma^\lambda\gamma^\rho 
\gamma_\mu (1-\gamma^5)v_u(q)
\int \iddk {(k-q)_\rho (k+q')_\sigma \over D }\ ,
\en
with the denominator
\be
D=(k^2-\mgd)_\Lambda (k+q')^2 (k-q)^2\ .
\en
Let us introduce the following notation for the
various integrals
\bea
&& \int\iddk {1\over D}= H(t) \ ,\nonumber\\
&& \int\iddk {k_\mu \over D}=H_0 (t)(q_\mu-q'_\mu) \ ,\nonumber\\
&& \int\iddk {k_\mu k_\nu\over D}=H_2(t)\, g_{\mu\nu}+ H_3(t)(q_\mu q_\nu
+q'_\mu q'_\nu)+H_4(t) (q_\mu q'_\nu +q_\nu q'_\mu) \ .
\ena
These integrals can be explicitly computed, yielding
\bea
&& H(t)=\facpi \left( {1\over t}\log{t\over\mgd}\log{t+\mgd\over\mgd}
+{1\over t}\dilog{t+\mgd\over\mgd} \right) \ ,\nonumber\\
&& H_0(t)=\facpi \left( -{1\over t} + {1\over t}\log {t\over\mgd}\right)
\ ,\nonumber\\
&& H_2(t)=\facpi \left( {1\over4}\,\div 
-{1\over4} \log{t\over\mu_0^2} +{3\over8}\right)
\ ,\nonumber\\
&& H_4(t)=\facpi \left(-{1\over 2t}\right)\ .
\ena
We note that $H_2(t)$ is the only integral which diverges 
as $\Lambda\to\infty$.
Simplifying the Dirac structure leads to an amplitude proportional 
to the leading order one,
\be
T_{23}=T_0\times (-Q_2 Q_3)\,  e^2
[4\,H_2 +2{t } (-H +2H_0 +H_4 ) ]\ .
\en
We keep only the terms which do not vanish when $\mg\to 0$, and we obtain
the decay width
\be
\Gamma_{23}=\Gamma_0\times Q_2 Q_3\, {\alpha\over2\pi}
\left[-\div +\log{\mld\over\mu_0^2}+4\LLdeps+{43\over3}\LLeps
+{859\over72}-{\pi^2\over3}\right]\ .
\en

\subsubsection{Contributions $Q_0Q_2$ and $Q_0Q_3$}

Here we considers the diagrams with one photon line joining the lepton
line to one of the quark lines. The contribution 
proportional to $Q_0Q_2$ (middle diagram in fig.~\ref{fig:oneloop}) 
is given by
\bea
T_{02}= &&\gff  Q_0Q_2 e^2\,
\int\iddk {1\over D_l}\times
\nonumber\\
&&\bar u_d(q')\gamma^\alpha (\kslash +\qqslash)\gamma^\lambda 
(1-\gamma^5) v_u(q)\,
\bar u_\nu(p')\gamma_\lambda(1-\gamma^5)[\kslash+\pslash+\ml]\gamma_\alpha
u_l(p)\ ,
\ena
with
\be
D_l=(k^2-\mgd)_\Lambda ( (k+p)^2-\mld)(k+q')^2\ .
\en
As in the previous case, we introduce the various Feynman integrals
\bea
&& \int\iddk {1\over D_l}= I(u)\nonumber \ ,\\
&& \int\iddk {k_\mu\over D_l}=I_0(u)p_\mu +I_1(u)q'_\mu\nonumber \ ,\\
&& \int\iddk {k_\mu k_\nu\over D_l}=I_2(u) g_{\mu\nu}
+I_3(u) p_\mu p_\nu +I_4(u) (p_\mu q'_\nu+p_\nu q'_\mu)+I_5(u) q'_\mu q'_\nu\ .
\ena
which can be computed easily
\bea
&&I(u) = 
\facpi {-1\over\mld-u}
\left[ \left( \LLeps-\log x_u\right)^2 + \dilog(x_u)
+{\pi^2\over4}\right]
\ ,\nonumber\\
&&I_0(u) =\facpi\left(-{1\over u}\right)\log x_u
\ ,\nonumber\\
&&I_1(u) =\facpi {1\over\mld- u}\left[ -2\LLeps-1 
+\left(1+{\mld\over u}\right)\log x_u\right]
\ ,\nonumber\\
&&I_2(u) =\facpi\left({1\over4}\right) \left[ 
  \div 
- \log{M_l^2\over\mu_0^2}
+ {\mld-u\over u}\log x_u 
+ {3\over2}\right]
\ ,\nonumber\\
&&I_3(u) =\facpi \left(-{1\over 2}\right)\left[ {1\over u} + 
{\mld-u\over u^2}\log x_u\right]
\ ,\nonumber\\
&&I_4(u) =\facpi \left({1\over 2}\right)\left[ {1\over u} + 
{\mld\over u^2}\log x_u\right]
\ .\ena
with
\be
x_u={\mld-u\over\mld}\ .
\en
All these integrals are convergent as $\Lambda\to\infty$ 
except $I_2$. Coming back to the amplitude,
we simplify the Dirac algebra and end up with
the following structure
\bea\lbl{T02}
&&T_{02}=-{G_F\over\sqrt2} Q_0Q_2 e^2   \Bigg\{ 
\bar u_d(q')\gamma^\nu(1-\gamma^5) v_u(q)\,
 \bar u_\nu(p')\gamma^\mu (1-\gamma^5) u_l(p)
\nonumber\\
&& \phantom{T_{02}=-{G_F\over\sqrt2} Q_0Q_2}
\quad\times 4[ I_2(u) g_{\mu\nu}+ p_\mu p_\nu (I_0(u)+I_3(u))
+q'_\mu p_\nu (I(u)+I_0(u)+I_1(u)+I_4(u))]
\nonumber\\
&&\phantom{T_{02}=-{G_F\over\sqrt2} Q_0Q_2}
+\bar u_d(q')\gamma^\lambda(1-\gamma^5)v_u(q)\, 
\bar u_\nu(p')\gamma^\mu  \gamma_\lambda (1+\gamma^5)u_l(p)
\nonumber\\
&&\phantom{T_{02}=-{G_F\over\sqrt2} Q_0Q_2}
\quad \times(-2M_l)[I_0(u) p_\mu +(I(u)+I_1(u)) q'_\mu]\Bigg\}\ .
\ena
The graph which gives the contribution proportional to $Q_0 Q_3$ 
is similar (right-hand diagram in fig.~\ref{fig:oneloop}), 
but it involves the functions $I_i(s)$ instead of
$I_i(u)$. In this case, the result reads
\bea\lbl{T03}
&&T_{03}=-{G_F\over\sqrt2} Q_0Q_3\, e^2   \Bigg\{ 
\bar u_d(q')\gamma^\lambda(1-\gamma^5) v_u(q)
 \, \bar u_\nu(p')\gamma_\lambda (1-\gamma^5) u_l(p)
\nonumber\\
&& \phantom{T_{02}=-{G_F\over\sqrt2} Q_0Q_2}
\quad\times 4[ 4\,I_2(s) + p^2 (I_0(s)+I_3(s)) +p.q(I(s)+I_0(s)+I_1(s)+2I_4(s))]
\nonumber\\
&&\phantom{T_{02}=-{G_F\over\sqrt2} Q_0Q_2}
+\bar u_d(q')\gamma^\lambda(1-\gamma^5)v_u(q)\,
\bar u_\nu(p')\gamma_\lambda \gamma^\mu  (1+\gamma^5)u_l(p)
\nonumber\\
&& \phantom{T_{02}=-{G_F\over\sqrt2} Q_0Q_2}
\quad \times(-2M_l)[I_0(s) p_\mu +(I(s)+I_1(s)) q_\mu]\Bigg\}\ .
\ena
One notices that the divergent piece in $T_{02}$ and in $T_{03}$ 
is proportional to the leading-order amplitude $T_0$.

The resulting one-loop corrections to the decay rate are
\bea
&&\Gamma_{02}=\Gamma_0 Q_0Q_2\, {\alpha\over 2\pi} 
\left[\div -\log{\mld\over\mu_0^2}
-2\LLdeps-{17\over3}\LLeps-{5\over6}\pi^2- {1\over8}\right]\ .
\\
&&\Gamma_{03}=\Gamma_0 Q_0Q_3\,{\alpha\over 2\pi} 
\left[4\,\div-4\log{\mld\over\mu_0^2}
-2\LLdeps-{19\over3}\LLeps-{5\over6}\pi^2+ {47\over72}\right]\ .
\nonumber
\ena

This completes the calculation of the electromagnetic corrections at order
$e^2$ to the decay amplitude~\rf{amplit} in Fermi theory. The result is
ultraviolet divergent as well as infrared divergent. The latter divergence 
disappears upon adding to the decay rate the one involving a real photon,
$\Gamma(l\to \bar u+d+\nu+\gamma)$ (its explicit expression
can be found in ref.~\cite{bl}). 
Ultraviolet divergences can be absorbed into local
counterterms which we discuss in the next section. 

\subsection{Counterterms and matching}

The previous calculations show
that the one-loop  ultraviolet divergences in the decay amplitude
$T(p,q,p',q')$ are proportional to the leading-order amplitude
$T_0(p,q,p',q')$. Thus, we may remove the divergences simply
by adding a set of four counterterms proportional to the original 
Fermi Lagrangian
\bea\lbl{ctlag}
&&{\cal L}_{CT}= \gffL e^2 \left\{
\psibl \gamma^\lambda \psinu  \times
\psibd \gamma_\lambda \psiu   +h.c. 
\right\}
\nonumber\\
&&\phantom{{\cal L}_{CT}= \gffL}
\times\left[ g_{00} Q_0^2 +g_{23} (Q_2+Q_3)^2
+g_{02} Q_0 Q_2 +g_{03}(-Q_0 Q_3) \right]\ .
\ena
We will recast some terms into a more standard form soon. At this stage however,
we just want to match the computation in the Fermi theory with that 
in the Standard Model. The decay amplitude can be made finite by imposing
the following relations among bare and renormalised couplings
in the Lagrangian~\rf{ctlag}
\bea
&& g_{00}=\mufacpi \left({1\over2}\,\divexp\right) + g^r_{00}(\mu_0)\nonumber
\ ,\\
&& g_{23}=\mufacpi \left({1\over2}\,\divexp\right) + g^r_{23}(\mu_0)\nonumber
\ ,\\
&& g_{02}=\mufacpi \left(-\,\divexp\right) + g^r_{02}(\mu_0)\nonumber \ ,\\
&& g_{03}=\mufacpi \left(  4\,\divexp\right) + g^r_{03}(\mu_0)\ ,
\ena
which leads to the renormalised one-loop correction to the
decay rate in Fermi theory
\bea\lbl{Geff}
&&\Gamma_{\rm Fermi}= e^2 \Gamma_0\, \Bigg\{ 
\facpi\left[ 
6(1+\Qbar)\log{\mu_0\over M_l}
+\Qbar\left({7\over4}+{3\over4}\Qbar\right)+{27\over2}
-2\pi^2  \right]
\\
&&\phantom{d\Gamma_{eff}=\Gamma_0 e^2\, \{  } 
+2g^r_{00}(\mu_0)+2 g^r_{23}(\mu_0) 
+(1-\Qbar) g^r_{02}(\mu_0) 
-(1+\Qbar) g^r_{03}(\mu_0)\Bigg\}\ ,
\nonumber
\ena
where the tree-level decay rate $\Gamma_0$ is given by eq.~\rf{Gamma0}.
The rate $\Gamma_{\rm Fermi}$ in eq.~\rf{Geff} also includes the process with
a real photon in the final state and, as a consequence, there is no infrared
divergence any more. In this expression, we have restricted the values of the electric charges 
(which were arbitrary up to now) to their physical values for
$Q_0$ and the sum $Q_2+Q_3=Q_0$, but we have left arbitrary the difference
\be
\Qbar= Q_2-Q_3\ .
\en

One must then equate this expression for 
the decay rate to the Standard Model one, 
which reads according to ref.~\cite{bl}
\bea\lbl{GammaSM0}
&& \Gamma_{SM}= {\bar{\alpha}^2 M_l^5 \over 384\pi s_W^4 M_W^4}\Big\{ 1
-2{\Pi_W(0)\over M_W^2} 
+{\alpha\over 2\pi} \Big[ 3[1-Q_0(Q_2-Q_3)]\log{M_Z\over M_l} 
\nonumber\\
&&\phantom{d\Gamma_{SM}= {\alpha^2 M_l^5 \over 384\pi s_W^4 M_W^4}\Big\{}
+\Big[ {7\over s_W^4}-{4\over s_W^2}\Big]\log c_W 
+{6\over s_W^2}-{1\over2}+\Big[{89\over24}-\pi^2\Big] Q_0^2
\nonumber\\
&&\phantom{d\Gamma_{SM}= {\alpha^2 M_l^5 \over 384\pi s_W^4 M_W^4}\Big\{}
-{43\over24}( Q_2^2+Q_3^2) +{29\over6} Q_0Q_2 
+{237\over36}Q_0Q_3-{61\over12} Q_2Q_3 \Big]\Big\}
\ena
where $\bar\alpha$ is the running QED coupling constant,
$c_W$ and $s_W$ denote the cosine and sine of the Weinberg angle
and $\Pi_W(0)$ is the $W$-propagator correction, renormalised on the
mass shell and evaluated at zero momentum.
This expression is valid provided that electric charge conservation holds, i.e.
$Q_0=Q_2+Q_3$. The cancellation of ultraviolet divergences imposes
that $Q_0^2=1$, while the difference $Q_2-Q_3$ may be kept as a free parameter.

Before matching the expressions~\rf{Geff} and~\rf{GammaSM0}, we 
must express the combination of Standard Model parameters $\bar{\alpha}/(s^2_W
M^2_W)$ in terms of the Fermi coupling $G_F$. 
This relation is obtained by matching the expressions for the muon 
lifetime ($Q_0=Q_2=-1$ and $Q_3=0$) in both theories~\cite{bl}, 
including radiative corrections at one loop. Doing so provides the relation
between the running QED coupling and the Fermi constant
\be
{\bar{\alpha}\over\sqrt{2} s_W^2 M_W^2}= {G_F\over\pi} 
\left[1 +{\Pi_W(0)\over M_W^2}
+{\alpha\over 2s_W^2\pi}\left( \log c_W\left( 2-{7\over 2s_W^2}\right)-3\right)
\right]\ .
\en
Replacing in eq.~\rf{GammaSM0} we can re-express BL's result in terms of
$G_F$,
\be\lbl{GammaSM}
\Gamma_{SM}=\Gamma_0\left\{1+ {\alpha\over4\pi} 
\left[ 6(1+\Qbar)\log{M_Z\over\ml}
+\Qbar \left({7\over4}+{3\over4}\Qbar\right)+{27\over2}-2\pi^2
\right]\right\}\ .
\en
Since $\Qbar$ need not be set to its physical value, 
matching eq.~\rf{GammaSM} with eq.~\rf{Geff} generates two 
independent equations for the counterterm coupling constants
\bea\lbl{sebrels}
&& g^r_{02}(\mu_0) + g^r_{03}(\mu_0)= 
\facpi\left[ -6 \log{M_Z\over\mu_0} \right]\ ,
\nonumber\\
&& g^r_{00}(\mu_0) + g^r_{23}(\mu_0) + g^r_{02}(\mu_0) =0\ .
\ena
This ends the first matching step.
\section{Matching Fermi theory and the chiral Lagrangian at one loop }
The second matching step proceeds in a rather different way from the 
first one. We will consider the effective chiral Lagrangian in which spurion
sources are introduced for the purpose of classifying the independent terms.
The trick will be to define correlators associated with these sources
and to compute them in the two different effective theories.
\subsection{Chiral Lagrangian with dynamical photons and leptons}
Coupling QCD to electromagnetism breaks chiral symmetry explicitly
because the quark charge matrix $Q$ is not proportional to the unit matrix. 
Coupling to the weak interaction generates an additional breaking
induced by the weak charge matrix $\Qw$. 
We will apply the chiral expansion to the three lightest quarks such
that these matrices are
\be
Q=\left(
\begin{array}{ccc}
{2\over3} &0&0\\
0&-{1\over3}&0\\
0&0&-{1\over3}
\end{array}\right),\quad
\Qw= -2\sqrt2  \left(
\begin{array}{ccc}
0&V_{ud}&V_{us}\\
0&0&0\\
0&0&0
\end{array}\right)\ .
\en
The neutral current part of the weak interaction will not be considered here.
At the level of the effective Lagrangian, the 
symmetry breaking induced by $Q$ and $\Qw$ 
can be accounted for and treated perturbatively by using
the spurion formalism. The treatment is analogous to the case of the symmetry 
breaking caused by the quark mass matrix ${\cal M}$. In that case, one
replaces the physical mass matrix by a pair of sources $s(x)$, $p(x)$ to which
one ascribes a transformation rule under the chiral group~\cite{gl84,gl85}
\be
s(x)+ip(x)\rightarrow g_R\, [ s(x)+ip(x)]\, g_L^\dagger\ ,
\en
where $(g_L,g_R)$ is a group element. 
In the same manner, one replaces the electric charge 
matrix $Q$ by two spurion sources
$\ql(x)$, $\qr(x)$~\cite{urech95} and the weak charge 
matrix $\Qw$ by one spurion source $\qw(x)$~\cite{knrt}. 
The part of the Lagrangian accounting for the coupling of the light quarks to
the photon and to a lepton pair is then written as
\bea\lbl{linspur}
&&{\cal L}^{spurions}_{QCD+Fermi}= 
-e\,F_\lambda 
   (\overline{\psi}_L\ql  \gamma^\lambda\psi_L + 
   \overline{\psi}_R\qr  \gamma^\lambda\psi_R) 
\nonumber\\
&&\phantom{{\cal L}^{spurions}_{QCD+Fermi}= -e\,F_\lambda }
-{4G_F\over\sqrt2}\left( \psibl \gamma_\lambda\psinu
   \,\overline{\psi}_L\qw   \gamma^\lambda\psi_L 
                        +\psibnu \gamma_\lambda\psil
   \,\overline{\psi}_L\qwdag\gamma^\lambda\psi_L \right)\ ,
\ena
where $\psi$ collects the $u$, $d$, $s$ quark fields. 
Chiral invariance is satisfied provided the spurion sources
are assumed to transform as
\be
\qr(x)\to g_R\, \qr(x)\,g_R^\dagger,\ 
\ql(x)\to g_L\, \ql(x)\,g_L^\dagger,\ 
\qw(x)\to g_L\, \qw(x)\,g_L^\dagger\ .
\en
It is also convenient to endow the spurions with the chiral order
\be\lbl{count1}
\ql,\ \qr,\ \qw\ \sim O(p^0)\ .
\en

Having defined the transformations of the spurion fields, one
can build the most general effective Lagrangian satisfying 
chiral symmetry with pseudo-Goldstone bosons, photon and light leptons
as dynamical fields. This Lagrangian provides a complete low-energy
description of the Standard Model. We deal with massless quarks, which means
that we take the chiral limit $m_u=m_d=m_s=0$ (in the following, QCD 
in this limit will be called ``chiral QCD'' for concision).

The pseudo-Goldstone mesons ($\pi$, $K$, $\eta$) are included into a 
unitary matrix $U=u^2$ and a so-called ``building block'' $u_\mu$
(see e.g.~\cite{egpr}) 
\be
u_\mu= iu^\dagger D_\mu U u^\dagger,\quad
D_\mu U = \partial_\mu U -ir_\mu U +i U l_\mu\ ,
\en
where $l_\mu$, $r_\mu$ contain not only vector and axial-vector sources for
the corresponding QCD currents, but also the photon, the light leptons 
and the spurion sources,
\bea
&&l_\mu=v_\mu-a_\mu -e \ql F_\mu + G_F [ \qw\psil\gamma_\mu\psinu +
          {\qwdag} \overline{\nu}_L\gamma_\mu l_L]\nonumber \ ,\\
&&r_\mu=v_\mu+a_\mu -e \qr F_\mu \ .
\ena
Chiral ``building blocks'' may be constructed from the spurion fields
\be 
\qqr\equiv u^\dagger \qr u,\ 
\qql\equiv u \ql u^\dagger,\ 
\qqw\equiv u \qw u^\dagger \ ,
\en
and also
\be
\qqrmu\equiv u^\dagger D^\mu \qr u,\ 
\qqlmu\equiv u D^\mu \ql u^\dagger\ . 
\en
The covariant derivatives for the spurions are defined as~\cite{urech95}
\be\lbl{derispur}
D^\mu \qr\equiv \partial^\mu \qr -i [r_\mu,\qr],\quad
D^\mu \ql\equiv \partial^\mu \ql -i [l_\mu,\ql]\  .
\en
In ref.~\cite{urech95}, the independent terms that contain one pair of 
spurions $\ql$, $\qr$ and that contribute up to $O(p^4)$ were classified.
One of these terms will play a special role in our discussion, namely
\be
{\cal L}^{(12)}_{\rm Urech}= 
-ie^2 F_0^2K_{12}\trace{ u_\mu ([\qqlmu,\qql]-[\qqrmu,\qqr])}\ .
\en
Knecht \emph{et  al.}~\cite{knrt} listed the $O(p^4)$ elements of the chiral
Lagrangian that involve a light lepton pair and are associated
with semi-leptonic decays. They obtained seven independent terms, once
the following constraint was implemented
\be\lbl{constq}
\ql \qw ={2\over3} \qw,\quad
\qw \ql =-{1\over3} \qw \ .
\en
Since we want to 
discuss the physical interpretation of the associated LEC's $X_i$,
it is convenient to consider a somewhat more general situation and
relax the constraint~\rf{constq}.
This leads to an extended chiral Lagrangian which contains two
additional terms. The associated LEC's will be called $\hat X_1$, $\hat X_2$. 
The remaining terms and the associated LEC's are identical to the case
considered in ref.~\cite{knrt}, except for the LEC's $X_6$ and $X_7$ which
have different values in the two settings and which will be labelled 
$\hat X_6$, $\hat X_7$ in our case.
The extended Lagrangian reads
\bea\lbl{chirlag}
&&{\cal L}_{\rm leptons}= e^2 \sum_l \Big\{ F_0^2 G_F\Big[
{X_1}\bar{l}_L\gamma_\mu\nu_L \trace{u^\mu\{\qqr,\qqw\} } 
+{\hat X_1}\bar{l}_L\gamma_\mu\nu_L \trace{u^\mu\{\qql,\qqw\} } 
\nonumber\\
&&\phantom{{\cal L}_{leptons}= e^2 \sum_l \Big\{ F_0^2 \Big[}
+{X_2}\bar{l}_L\gamma_\mu\nu_L\trace{u^\mu[\qqr,\qqw]}
+{\hat X_2}\bar{l}_L\gamma_\mu\nu_L\trace{u^\mu[\qql,\qqw]}
\nonumber\\
&&\phantom{{\cal L}_{leptons}= e^2 \sum_l \Big\{ F_0^2 \Big[}
+{X_3}\ml\bar{l}_R \nu_L\trace{\qqr\qqw}
\nonumber\\
&&\phantom{{\cal L}_{leptons}= e^2 \sum_l \Big\{ F_0^2 \Big[}
+i{X_4}\bar{l}_L\gamma_\mu\nu_L\trace{\qqlmu \qqw }
+i{X_5}\bar{l}_L\gamma_\mu\nu_L\trace{\qqrmu \qqw} +h.c.\Big]
\nonumber\\
&&\phantom{{\cal L}_{leptons}= e^2 \sum_l \Big\{ F_0^2 \Big[}
+ {\hat X_6}\bar l (i\dslash+e\Aslash)l 
+ {\hat X_7} M_l\bar{l}l\Big\}\ .
\ena
An additional possible term of the form $\ml\bar{l}_R \nu_L\trace{\ql\qw}$
is of no practical relevance and will be ignored.
The original LEC's $X_6$ and $X_7$ are easy to relate to the new set of LEC's $\hat X_i$
\bea\lbl{x6x7}
&& X_6= \hat X_6+{4\over3} \hat X_1 +4\hat X_2\nonumber \ ,\\
&& X_7= \hat X_7-{4\over3} \hat X_1 -4\hat X_2\ .
\ena
These relations allow us to disentangle the  strong-interaction content
of $X_6$ and $X_7$, corresponding to $\hat X_1$ and $\hat X_2$, and the
electroweak contributions encoded in $\hat X_6$ and $\hat X_7$.

Finally, let us make two remarks. Firstly, one can
verify that the Lagrangian terms listed in~\rf{chirlag} do have chiral order
$p^4$ if one uses the counting rules~\rf{count0} and \rf{count1}.
Secondly,  eq.~\rf{chirlag} obviously does not exhaust the possible terms 
of order $p^4$ involving light lepton pairs: they include only those connected
with charged currents. Terms related with neutral currents are disregarded
here (for some examples, see e.g.~\cite{savage}).

\subsection{Spurion correlators}\label{sec:chspur}

We have included electric and weak charge spurion sources in the
Lagrangian. Therefore, in addition to the usual vector, 
axial-vector{\ldots} sources, the generating functional depends
on $\ql(x)$, $\qr(x)$, $\qw(x)$. We can define generalised Green 
functions by taking derivatives of the generating functional 
with respect to these sources,
and eventually with respect to the usual sources, in order to 
compute matrix elements between physical states. This idea was used
in ref.~\cite{bm97} to generate a set of sum rules
obeyed by the LEC's $K_i$~\cite{urech95}. It can be extended to the present
situation without difficulty, and we define
a set of three matrix elements of three operators, obtained by taking
one functional derivative with respect to an electric charge 
spurion and one derivative with respect to a weak charge spurion. 

More specifically, we introduce the 
charge spurions $\qv(x)$ and $\qa(x)$ as
\be
\ql(x)= {1\over2} (\qv(x)-\qa(x)),\quad
\qr(x)= {1\over2} (\qv(x)+\qa(x))\ .
\en
and the correlators
\be\lbl{gcorr}
i\,\int d^4x\,{\rm e}^{irx}
\trace{ l(p) \bar\nu(q)\vert 
{\delta^2 W(\ql, \qr, \qw)\over \delta \qrb(x) \delta \qwc(0)}
 \vert 0}
\equiv \delta^{bc} G_{RW}(p,q,r)\ .
\en
and
\bea\lbl{fdcorr}
&& \int d^4x\, \trace{ l(p) \bar\nu(q)\vert 
{\delta^2 W(\qv, \qa, \qw)\over \delta \qvb(x) \delta \qwc(0)}
\vert \pi^a(r)}
\equiv if^{abc}F_{VW}(p,q,r)  + d^{abc} D_{VW}(p,q,r) 
\nonumber\\
&& \int d^4x\, \trace{ l(p) \bar\nu(q)\vert 
{\delta^2 W(\qv, \qa, \qw)\over \delta \qab(x) \delta \qwc(0)}
\vert \pi^a(r)}\equiv if^{abc} F_{AW}(p,q,r)  + d^{abc} D_{AW}(p,q,r)  \ .
\ena
where $f_{abc}$ and $d_{abc}$ denote the standard antisymmetric and symmetric
functions defined through the commutation and anticommutation of
Gell-Mann matrices. Once the functional derivatives 
have been taken, we set all sources to zero
(including the charge spurions).

In the following, we will compute these generalised
correlators in two different ways: firstly
from the chiral Lagrangian, leading to expressions in terms
of low-energy coupling constants, and secondly from the QCD and
Fermi Lagrangians, yielding the correlators in terms of the counterterms in
Fermi theory. This approach allows one to generate
representations of the chiral coupling constants
in terms of pure QCD correlation functions in a rather 
straightforward way, with a clear identification of
the short-distance contributions from the Standard Model.
 
\subsection{Correlators from the chiral Lagrangian at one loop}\label{sec:chir}

Let us start with the chiral Lagrangian. The spurion correlators receive
tree-level contributions from $O(p^4)$ LEC's, and 
one-loop contributions with $O(p^2)$ vertices.
Let us illustrate this in the case of $G_{RW}(p,q,r)$. 
The tree contribution involves $X_3$ and $X_5$
\be\lbl{GRWtree}
G^{\rm tree}_{RW}(p,q,r)= {1\over2}e^2 G_F F^2_0 \Big[ 
M_l X_3 \bar u_l(p){1-\gamma_5\over2} v_\nu(q) 
-X_5 u_l(p)\gamma_\mu {1-\gamma_5\over2} v_\nu(q)r^\mu \Big]\ .
\en
The one-loop contribution has the following expression
\bea\lbl{gwloop}
&& G^{\rm loop}_{RW}= - {e^2 Q_0 G_F F_0^2\over 4}\int\iddkd \left[
{(k+r)^\sigma (k+r)^\lambda\over (k+r)^2} -g^{\sigma\lambda}\right]\\
&&\phantom{  G^{loop}_{RW}= - {e^2 Q_0 G_F} }
{1\over (k^2-\mgd)( (k-p)^2-\mld)}\ubarl \gamma_\sigma (\pslash -\kslash
+\ml)\gamma_\lambda \nuL\nonumber\ .
\ena
In this sector, the ultraviolet divergences will be controlled via dimensional
regularisation (as usual in chiral perturbation theory).
We must compute $G^{\rm loop}_{RW}$ only up to $O(r)$. This means
that we may expand the integral for small values of the pion momentum $r$ up
to linear order. A further simplification consists in expanding
in powers of the lepton mass $\ml$ around the limit $M_l=0$, keeping 
$M_\gamma\ne0$ whenever necessary to avoid infrared divergences.
After performing these expansions, the loop contribution exhibits the
following explicit expression
\bea\lbl{GRWloop}
&& G^{\rm loop}_{RW}(p,q,r)= - {e^2 Q_0 G_F F_0^2}\Bigg\{ \nonumber\\
&&\phantom{G^{loop}_{RW}=}
M_l\ubarl\nuL  
\left[{3\over2}\divchi +{1\over16\pi^2}\left(
      {3\over4}\log{\mgd\over\mud}+{1\over8}\right)\right]
\nonumber\\
&&\phantom{G^{loop}_{RW}=}
+\ubarl\gamma_\mu\nuL r^\mu
\left[{3\over4}\divchi +{1\over16\pi^2}\left(
      {3\over8}\log{\mgd\over\mud}+{1\over16}\right)\right]\Bigg\}\ ,
\ena
where the chiral divergence is defined in the customary way
\be
\divchi ={\mu^{d-4}\over16\pi^2}\left\{
{1\over d-4} -{1\over2}\left( \log4\pi-\gamma+1\right)\right\}\ .
\nonumber
\en
The complete chiral expression for $G_{RW}$ is obtained by adding
tree~\rf{GRWtree} and one-loop~\rf{GRWloop} pieces 
\be \lbl{Gsumchiral}
G_{RW}^{\rm chir}(p,q,r)=G_{RW}^{\rm tree}(p,q,r)+G_{RW}^{\rm loop}(p,q,r)\ .
\en
The ultraviolet divergences are absorbed into the LEC's
\be
X_i= X_i^r(\mu) +\Xi_i\,{\mu^{d-4}\over16\pi^2}\left(
{1\over d-4} -{1\over2} ( \log4\pi-\gamma+1) \right)\ .
\en
This requirement sets the divergence coefficients
\be
\Xi_3=-3\ ,\qquad \Xi_5= {3\over2}\ ,
\en
in agreement with ref.~\cite{knrt}. We proceed in exactly the same way 
with the other spurion correlators $F_{VW}$, $D_{VW}$, $F_{AW}$, $D_{AW}$ 
(the loop contribution to $D$-terms involve correlators that vanish by
invariance under charge conjugation). These correlators, 
expanded up to linear order in the momentum $r$, 
have the following expressions at next-to-leading order
\bea\lbl{treemat}
&& F^{\rm chir}_{VW}(p,q,r)= F^{\rm chir}_{VW}(p,q,0)
+ e^2 G_F F_0\, \bar u_l(p)\gamma_\mu {1-\gamma_5\over2} v_\nu(q)r^\mu
\nonumber\ \\
&&\phantom{ F^{\rm chir}_{VW}(p,q,r)= F^{\rm chir}_{VW}(p,q,0)}
\times\left[ X^r_2+\hat X^r_2 +\facpi \left( {5\over4}\log{\mgd\over\mu^2}
+{1\over8}\right) \right]\nonumber\ ,\\
&& F^{\rm chir}_{AW}(p,q,r)= F^{\rm chir}_{AW}(p,q,0)
+ e^2 G_F F_0\, \bar u_l(p)\gamma_\mu {1-\gamma_5\over2} v_\nu(q)r^\mu
\nonumber\ \\
&&\phantom{ F^{\rm chir}_{VW}(p,q,r)= F^{\rm chir}_{VW}(p,q,0)}
\times\left[ X^r_2-\hat X^r_2 +\facpi \left(-{1\over2}\log{\mgd\over\mu^2}
\right)\right]\nonumber\ ,\\
&& D^{\rm chir}_{VW}(p,q,r)=
e^2 G_F F_0\,\bar u_l(p)\gamma_\mu {1-\gamma_5\over2} v_\nu(q)r^\mu
[X_1+\hat X_1]\nonumber\ ,\\
&& D^{\rm chir}_{AW}(p,q,r)=
e^2 G_F F_0\,\bar u_l(p)\gamma_\mu {1-\gamma_5\over2} v_\nu(q)r^\mu
[X_1-\hat X_1]\ .
\ena
We have not written the explicit formulas for $F^{\rm chir}_{VW}(p,q,0)$
and $F^{\rm chir}_{AW}(p,q,0)$. The following simple relation holds
\be \lbl{softpion}
F^{\rm chir}_{VW}(p,q,0)=F^{\rm chir}_{AW}(p,q,0)=G^{\rm chir}_{RW}(p,q)\ ,
\en
as a result of a soft-pion theorem  (see eq.~\rf{eqdmo} below). 
The coefficients of the chiral divergences are
\be
\Xi_1=\hat\Xi_1=0\ ,\qquad \Xi_2 =-{3\over4}\ ,\qquad \hat\Xi_2=-{7\over4}\ ,
\en
also in agreement with ref.~\cite{knrt}.

\subsection{Correlators from QCD + Fermi theory}
Here we compute the correlators introduced 
in sec.~\ref{sec:chspur} using the QCD and Fermi
Lagrangians. A first (non-local) contribution stems from
the terms in these Lagrangians which are linear in the spurion sources,
see eq.~\rf{linspur}. A second (local) contribution is due to
the counterterms in Fermi theory that are quadratic in the spurions.

\subsubsection{Integral contributions}\label{sec:int}

Let us first consider the contribution to the spurion correlators
coming from the Lagrangian eq.~\rf{linspur}. Matrix
elements of vector and axial-vector currents appear by
taking the functional derivatives defining the spurion correlators.
Let us introduce the following notation for these objects,
\bea\lbl{corrdef}
&&i\int d^4x\, {\rm e}^{ikx} \langle 0\vert V_\sigma^b(x) V_\lambda^c(0)-
A_\sigma^b(x) A_\lambda^c(0)\vert 0\rangle \equiv \delta^{bc}
\Pi^{\sigma\lambda}_{VV-AA}(k)
\nonumber\ ,\\
&&\int d^4x\,{\rm e}^{ikx}
\langle 0\vert V^b_\sigma(x)  V^c_\lambda(0)\vert \pi^a(r) \rangle \equiv
d^{abc} \Gamma^{\sigma\lambda}_{VV} (k,r)\nonumber\ ,\\
&&\int d^4x\,{\rm e}^{ikx}
\langle 0\vert A^b_\sigma(x)  A^c_\lambda(0)\vert \pi^a(r) \rangle \equiv
d^{abc} \Gamma^{\sigma\lambda}_{AA} (k,r)\nonumber\ ,\\
&&\int d^4x\,{\rm e}^{ikx}
\langle 0\vert V^b_\sigma(x)  A^c_\lambda(0)\vert \pi^a(r) \rangle \equiv
if^{abc} \Gamma^{\sigma\lambda}_{VA} (k,r) \ .
\ena
The choice between the $f^{abc}$ and the $d^{abc}$ tensor
in these equations is dictated by invariance under charge conjugation.
Let us remark that the Pauli-Villars regularisation offers a very appealing
feature here: the operators and matrix elements appearing in eqs.~\rf{corrdef} are not
to be defined in an arbitrary number of dimensions, but only in the physical
(four-dimensional) case.
The Lagrangian eq.~\rf{linspur} leads to contributions to the spurion
correlators that are integrals involving the QCD Green
function and vertex operators introduced above~\rf{corrdef}.
\bea\lbl{vacint}
&&G_{RW}^{\rm int}(p,q,r)= - {e^2 Q_0 G_F \over 4}\int \iddk 
\Pi^{\sigma\lambda}_{VV-AA}(k+r)\times K_{\sigma\lambda}(k,p,q)\ ,
\ena
with
\be
K_{\sigma\lambda}(k,p,q)={1\over (k^2-\mgd)_\Lambda( (k-p)^2-\mld)}
\ubarl \gamma_\sigma (\pslash-\kslash+\ml)\gamma_\lambda \nuL \ ,
\en
and
\bea\lbl{intrep}
&& F^{\rm int}_{VW}(p,q,r)= {e^2Q_0 G_F\over 2}
\int \iddk \Gamma^{\sigma\lambda}_{VA}(k,r)\times K_{\sigma\lambda}(k,p,q)
\nonumber\ ,\\
&& D^{\rm int}_{VW}(p,q,r)= -{e^2Q_0 G_F\over 2}\int \iddk \Gamma^{\sigma\lambda}_{VV}(k,r) \times K_{\sigma\lambda}(k,p,q)
\nonumber\ ,\\
&& F^{\rm int}_{AW}(p,q,r)= {e^2Q_0 G_F\over 2}\int \iddk \Gamma^{\lambda\sigma}_{VA}(r-k,r) \times K_{\sigma\lambda}(k,p,q)
\nonumber\ ,\\
&& D^{\rm int}_{AW}(p,q,r)= {e^2Q_0 G_F\over 2}\int \iddk \Gamma^{\sigma\lambda}_{AA}(k,r) \times K_{\sigma\lambda}(k,p,q)\ .
\ena
The integral in eq.~\rf{vacint} converges when the Pauli-Villars regulator
mass $\Lambda$ is sent to infinity (there is no ultraviolet divergence), whereas
the other integrals would diverge.
The divergences will be removed upon adding the contributions generated from
the Fermi counterterms.

\subsubsection{Counterterm contributions}
In order to identify the contributions to the spurion correlators 
arising from the Fermi counterterms~\rf{ctlag}, we must first
rewrite the Lagrangian in terms of spurion sources. After some 
manipulations, we can re-express the counterterms as follows
\bea\lbl{ctlag2}
&& {\cal L}_{\rm CT}= 
 -2 e^2 Q_0^2 g_{00}\, \overline{l} (i\dslash -eQ_0\Aslash-M_l) l
\nonumber\\
&&\phantom{{\cal L}_{CT}=}
- i e^2 g_{23}\, \Big( 
\overline{\psi}_L [\ql,D^\mu \ql]\gamma_\mu \psi_L +L\leftrightarrow R\Big)
\\
&&\phantom{{\cal L}_{CT}=}
 + {e^2Q_0G_F } 
\Bigg\{             \psibl \gamma_\lambda  \psinu\times  
\Big[
 g_{02}\, \overline{\psi}_L \gamma^\lambda\,  \qw \ql\psi_L 
+g_{03}\, \overline{\psi}_L \gamma^\lambda\,  \ql \qw\psi_L   \Big] +h.c. 
\Bigg\}\ .
\nonumber
\ena
The term proportional to $g_{00}$ has been written in a more conventional
way, which is equivalent to the formulation
in eq.~\rf{ctlag} as far as the amplitude $T$ is concerned (we have applied
equations of motion).
We have extended the term proportional to $g_{23}$ to comply with
the transformation laws of the spurions: this extended term contains
the piece proportional to $g_{23}$ in the original
formulation~\rf{ctlag}, as can be seen from the definition
of the spurion derivative~\rf{derispur}. Modulo these transformations, it is simple to check 
that setting the spurions to the physical charges $\ql=\qr=Q$, $\qw=\Qw$
reproduces the Lagrangian eq.~\rf{ctlag}. Up to terms which are physically
irrelevant, the translation from charge labels to spurions is essentially
unique.

In this new form, it is an easy task to compute the functional derivatives and deduce
the contributions to the spurion correlators. The following results are
obtained
\bea\lbl{ctmat}
&& G^{\rm CT}_{RW}=0\nonumber \ ,\\
&& F^{\rm CT}_{VW}= e^2 G_F Q_0
F_0\bar u_l(p)\gamma_\mu {1-\gamma_5\over2} v_\nu(q)r^\mu
\,\left[{1\over4}g_{02} - {1\over4} g_{03}\right]\nonumber \ ,\\
&& D^{\rm CT}_{VW}= e^2 G_F Q_0
F_0\bar u_l(p)\gamma_\mu {1-\gamma_5\over2} v_\nu(q)r^\mu
\,\left[-{1\over4}g_{02} - {1\over4} g_{03}\right]\nonumber \ ,\\
&& F^{\rm CT}_{AW}= e^2 G_F Q_0
F_0\bar u_l(p)\gamma_\mu {1-\gamma_5\over2} v_\nu(q)r^\mu
\,\left[-{1\over4}g_{02} + {1\over4} g_{03}\right]\nonumber \ ,\\
&& D^{\rm CT}_{AW}= e^2 G_F Q_0
F_0\bar u_l(p)\gamma_\mu {1-\gamma_5\over2} v_\nu(q)r^\mu
\,\left[+{1\over4}g_{02} + {1\over4} g_{03}\right] \ .
\ena
We can now add these contributions to the integral contributions
\be \lbl{Gsumfermi}
G^{\rm Fermi}_{RW}(p,q,r)=G^{\rm CT}_{RW}(p,q,r)+G^{\rm int}_{RW}(p,q,r)
\en
and similarly for the other correlators.
The result should be finite as $\Lambda\to\infty$: we verify this now
and show that the integrals can be brought to fairly simple forms. 

\subsection{Explicit representations of the chiral coupling 
constants}\label{sec:srules}

\subsubsection{Integral representations}

We can match the two expressions for the spurion correlators: the integral 
representation stemming from Fermi theory, such as eq.~\rf{Gsumfermi}, and the formulae 
obtained from the chiral effective Lagrangian, see eq.~\rf{Gsumchiral}. To do so,
let us expand the integral representations discussed in sec.~\ref{sec:int}
for small values of the pion momentum $r$, and compare the series with the 
chiral expansion derived in sec.~\ref{sec:chir}.
This comparison yields integral representations for the LEC's of the chiral
Lagrangian, which can be simplified further by displaying the kinematical
structures and the associated form factors of the correlators involved. Let us first introduce
the correlators related to $G_{RW}$, $D_{VW}$ and $D_{AW}$
\bea\lbl{ffactor0}
&&\Pi^{\rho\sigma}_{VV-AA}(k)\equiv F_0^2
(k^\rho k^\sigma-k^2 g^{\rho\sigma})\,\Pi_{VV-AA}(k^2)
\nonumber \ ,\\
&&\Gamma^{\rho\sigma}_{VV}(k,r)=iF_0
\epsilon^{\rho\sigma\alpha\beta} k_\alpha r_\beta \,\Gamma_{VV}(k^2,k.r)
\nonumber \ ,\\
&&\Gamma^{\rho\sigma}_{AA}(k,r)=iF_0
\epsilon^{\rho\sigma\alpha\beta} k_\alpha r_\beta \,\Gamma_{AA}(k^2,k.r)\ .
\ena
In practice, we need $\Gamma^{\rho\sigma}_{VV}$ and $\Gamma^{\rho\sigma}_{AA}$
only up to $O(r)$, and thus it is enough to get the form factors $\Gamma_{VV}(k^2,k.r)$ and 
$\Gamma_{AA}(k^2,k.r)$ in the limit where the pion momentum $r$ is set to zero.
We use the simplified notation
\be
\lim_{r\to0}\Gamma_{VV}(k^2,k.r)\equiv \Gamma_{VV}(k^2)\ ,\qquad
\lim_{r\to0}\Gamma_{AA}(k^2,k.r)\equiv \Gamma_{AA}(k^2)\ .
\en
Then, in connection with the spurion correlators $G_{RW}$, $D_{VW}$ and
$D_{AW}$, we can obtain representations for the four LEC's 
$X_1$, $\hat X_1$, $X_3$ and $X_5$
\bea\lbl{explixi}
&& X_1 = -\, {3\over8}\int\iddk\, {1\over k^2} (\Gamma_{VV}(k^2)
-\Gamma_{AA}(k^2))
\nonumber\ ,\\
&&\hat X_1=-\,{3\over8}\int\iddk\, {1\over k^2 }
\left( \Gamma_{VV}(k^2)+ \Gamma_{AA}(k^2) -{2\over k^2-\mu_1^2} \right) 
+{3\over4}\log{\mu_1^2\over M_Z^2}
\nonumber\ ,\\
&& X^r_3(\mu) = -\,{3\over2}\int\iddk\, 
{1\over k^2}\left( \Pi_{VV-AA}(k^2)+{\mu_1^2\over k^2 (k^2-\mu_1^2)}\right)
+\facpi\left(   {3\over2}\log{\mu^2\over\mu_1^2}-{1\over4}\right)
\nonumber\ ,\\
&& X^r_5(\mu) =  \,{3\over4}\int\iddk\, 
{1\over k^2}\left( \Pi_{VV-AA}(k^2)+{\mu_1^2\over k^2 (k^2-\mu_1^2)}\right)
\nonumber\\
&&\phantom{X^r_5(\mu) =  \,{3\over4}\int\iddk\,{1\over k^2}
\Pi_{VV-AA}(k^2)+}  
+\facpi\left(  -{3\over4}\log{\mu^2\over\mu_1^2}-{5\over8}\right)\ .
\ena
These integrals could be rewritten as one-dimensional integrals. 
In order to derive the expression of $\hat X_1$, we have re-expressed
the combination of counterterms $g^r_{02}(\mu_0)
+g^r_{03}(\mu_0)$ using the matching conditions eqs.~\rf{sebrels}. 
The result involves an explicitly convergent integral, 
as can be checked easily using the asymptotic behaviour 
(see e.g. \cite{knyff01}) 
of $\Gamma_{VV}(k^2)$, $\Gamma_{AA}(k^2)$, 
\be
\Gamma_{VV}(k^2),\ \Gamma_{AA}(k^2)\sim{1\over k^2} \ ,
\en 
still ignoring (for the moment) perturbative QCD corrections.
The scale $\mu_0$, related to the renormalisation in Fermi theory, 
has disappeared, which
signals that the original divergence was correctly cancelled by the
counterterm. An arbitrary scale $\mu_1$ has been introduced in the integrand 
to obtain convergent integrals, but the dependence on $\mu_1$ cancels
in the final result. 

The integrals involved in $X_1$, $X_2$, $X_5$   converge
because of the short-distance smoothness of the difference
$VV-AA$ in chiral QCD (see e.g.~\cite{weinbook}). In the case of  
$X_3$ and $X_5$, the integrands have been recast in a form which is explicitly
infrared finite. As in the previous case, the overall results are independent of the scale $\mu_1$
introduced in the integrands.

Similar sum rules can be written for $X_2$ and $\hat X_2$ by focusing on
the $r$-linear piece in $F_{AW}$ and $F_{VW}$. The function of interest is
the vertex correlator $\Gamma^{\sigma\lambda}_{VA}(p,r)$ which 
involves two form factors $F$ and $G$ in the chiral limit~\cite{bm97},
\be\lbl{GVA}
\Gamma^{\sigma\lambda}_{VA}(p,r)= F_0 \left\{ 
{(p^\sigma +2q^\sigma) q^\lambda\over q^2} -g^{\sigma\lambda}
+ F(p^2,q^2)\,P^{\sigma\lambda}
+ G(p^2,q^2)\,Q^{\sigma\lambda} \right\}\ ,
\en
with $q=r-p$ and
\be
P^{\sigma\lambda}=q^\sigma p^\lambda  -(p.q)\, g^{\sigma\lambda},\quad
Q^{\sigma\lambda}=p^2 q^\sigma q^\lambda 
                 +q^2 p^\sigma p^\lambda 
                - (p.q)\, p^\sigma q^\lambda
                - p^2 q^2 g^{\sigma\lambda}\ .
\en
In order to identify the LEC's $X_2$ and $\hat X_2$ one must 
expand $\Gamma^{\sigma\lambda}_{VA}(p,r)$ in eqs.~\rf{intrep} up to linear order
in the pion momentum $r$. Let us introduce
\bea
&&f(k^2)\equiv F(k^2,k^2),\ 
f_1(k^2)\equiv \partial_x F(x,k^2)\vert_{x=k^2},\ 
f_2(k^2)\equiv \partial_y F(k^2,y)\vert_{y=k^2}
\nonumber\\
&&g(k^2)\equiv G(k^2,k^2),\ 
g_1(k^2)\equiv \partial_x G(x,k^2)\vert_{x=k^2},\ 
g_2(k^2)\equiv \partial_y G(k^2,y)\vert_{y=k^2}
\ena
The correct QCD asymptotic behaviour of $\Gamma^{\sigma\lambda}_{VA}(k,r)$
as $k\to\infty$ (see~\cite{bm97}) is reproduced up to order $1/k^2$
provided that these functions obey the limits
\be\lbl{asylim}
\lim_{k^2\to\infty} k^4 g(k^2)= -1,\ 
\lim_{k^2\to\infty} k^4 f(k^2)= {\rm const.}\ ,\ 
\lim_{k^2\to\infty} k^4 (f_2(k^2)-k^2 g_2(k^2))=-{3\over2}\ .
\en
After some quick algebra, we find the following 
integral representations for $X_2$ and $\hat X_2$
(once again essentially one-dimensional)
\bea\lbl{explix2}
&& X_2^r(\mu)= 
-{3\over 8}\int\iddk {1\over k^2}
\left(\Pi_{VV-AA}(k^2)+ {\mu_1^2\over k^2(k^2-\mu_1^2)}\right)
+\facpi\left( {3\over8}\log{\mu^2\over\mu_1^2} +{5\over16}\right)
\nonumber\ ,\\
&&\hat X_2^r(\mu)= 
-{3\over8}\int\iddk \left[ {-1\over k^2(k^2-\mu_1^2)} +f_1(k^2)-f_2(k^2)
+k^2\,(-g_1(k^2)+g_2(k^2))\right]\nonumber\\
&& \phantom{X_2^r(\mu)=}
+\facpi\left( -{5\over4}\log{\mu_0^2\over\mu_1^2} 
              +{7\over8}\log{\mu^2\over\mu_1^2}
-{1\over16}\right) -{1\over4} g_{02}^r(\mu_0) + {1\over4} g_{03}^r(\mu_0)\ .
\ena
In order to derive these expressions we have used integration by parts, noting
that $f_1+f_2=f'$ and $g_1+g_2=g'$, as well as the soft-pion theorem~\cite{dmo}
\be\lbl{eqdmo}
\Gamma_{VA}^{\sigma\lambda}(k,0)={1\over F_0} 
\Pi^{\sigma\lambda}_{VV-AA}(k)\ .
\en
which implies that
\be
\Pi_{VV-AA}(k^2)=\frac{1}{k^2}-f(k^2)+k^2 g(k^2)\ .
\en
Let us remark that this soft-pion theorem, in combination with
eq.~\rf{softpion}, implies that the $O(r^0)$ pieces in $F_{AW}$ and $F_{VW}$
yield exactly the same sum rules as $G_{RW}$. 
One easily checks that the integral appearing in the expression of 
$\hat X_2^r$ is convergent whenever the form factors satisfy the
QCD asymptotic constraints eqs.~\rf{asylim}.
Remarkably, the LEC $X_2$ turns out to depend only on the Green
function $\langle VV-AA \rangle$. As in eq.~\rf{explixi}, 
a scale $\mu_1$ was introduced but the result is independent of $\mu_1$.
The result can also be verified to be independent of the scale $\mu_0$ which is a consequence
of the fact that the contribution from the counterterms correctly cancels 
the original divergence of the integral.

These exact integral representations reveal relationships among the coupling
constants which were not a priori expected
\bea
&& X^r_3(\mu)= 4 X^r_2(\mu)-{3\over2}\,\facpi\nonumber\ ,\\
&& X^r_5(\mu)=-2 X^r_2(\mu)\ .
\ena
Let us emphasise that these relations are absolutely general. In particular,
their validity is completely independent of any particular model for the
two- and three-point Green functions involved in the integral representations.

\subsubsection{The case of $X_6$}\label{sec:secx6}

Among the LEC's which are physically relevant, $X_6$ plays a special role. 
According to eq.~\rf{x6x7}, $X_6$ can
be expressed in terms of $\hat X_1$ and $\hat X_2$, which were discussed above,
and $\hat X_6$. By construction, $\hat X_6$ has no strong-interaction content:
it can be determined by computing the lepton wave-function renormalisation
factor $K_F$ in chiral perturbation theory and identifying it with our calculation
in Fermi Theory in sec.\ref{sec:self}. The regularisation schemes are different: the former 
employs dimensional regularisation and chiral
$\overline{MS}$ renormalisation, whereas the latter relies on Pauli-Villars
regularisation. We get the relation
\be
\hat X_6(\mu_0)= -2\,g^r_{00}(\mu_0) +{3\over2}{1\over 16\pi^2}\ .
\en
The resulting expression for $X_6$ involves a combination 
of counterterms, $-2g_{00}-g_{02}+g_{03}$, which is not
determined by the matching conditions~\rf{sebrels}. 
This implies that physical quantities must involve $X_6$ together with one 
additional, electromagnetic, LEC. It is not difficult to see that this 
LEC must be  $K_{12}$. We will see that the physically relevant combination is
\be\lbl{x6eff}
X_6^{\rm phys}(\mu)\equiv X_6^r(\mu)-4 K_{12}^r(\mu)  
=4\,( \hat X_2^r(\mu)-K_{12}^r(\mu)) + \hat X_6^r(\mu)
+{4\over3} \hat X_1^r(\mu)\ .
\en
The LEC $K_{12}$ was shown to satisfy an integral representation
in terms of the vertex function $\Gamma^{\sigma\lambda}_{VA}(k,r)$ that
we have introduced above~\cite{bm97}. Using the 
present notation and regularisation 
scheme, one can derive the explicit representation
\bea\lbl{explik12}
&& K_{12}^r(\mu)= -{3\over8}\int\iddk \left[ 
{-1\over k^2(k^2-\mu_1^2)} +f_1(k^2)-f_2(k^2)+k^2\,(-g_1(k^2)+g_2(k^2))\right]
\nonumber\\
&&\phantom{K_{12}^r(\mu)= }
+\facpi\left( -{1\over4} \log{\mu_0^2\over\mu_1^2} 
               -{1\over8}\log{\mu^2  \over\mu_1^2}
-{5\over16} \right) +{1\over2}\, g^r_{23}(\mu_0)
\ena
where integration by parts was used to simplify the formula. 

Let us now consider the difference $\hat X_{2}-K^r_{12}$ . 
Using the integral expressions~\rf{explik12} and~\rf{explix2},
one observes that all terms cancel except for the counterterms
\be\lbl{diff12}
\hat X^r_2(\mu)- K^r_{12}(\mu)=
-{1\over4} g_{02}^r(\mu_0) + {1\over4} g_{03}^r(\mu_0)
 - {1\over2}\, g^r_{23}(\mu_0) +\facpi\left( -\log{\mu_0^2\over\mu^2}
+{1\over4}\right)\ .
\en
Inserting this result into $X_6^{\rm phys}$ and setting $\mu_0=\mu$,
one realises that the resulting combination of counterterms is indeed 
determined from the matching conditions~\rf{sebrels}
\be
-g^r_{02}(\mu)+g^r_{03}(\mu)-2g^r_{00}(\mu)-2g^r_{23}(\mu)=\facpi\left(-6
\log{M_Z\over\mu}\right)\ .
\en
One ends up with the following simple representation of $X_6^{\rm phys}$
\bea\lbl{x6int}
&&X_6^r(\mu)-4K_{12}^r(\mu)=- {1\over2}\int\iddk {1\over k^2}
\left( \Gamma_{VV}(k^2)+\Gamma_{AA}(k^2)-{2\over k^2-\mu_1^2}\right)
\nonumber\\
&&\phantom{X_6^r(\mu)-4K_{12}^r(\mu)}
+\facpi\left[ -8\log{M_Z\over \mu_1} +3\log{\mu^2\over \mu_1^2} +{5\over2}
\right]\ .
\ena
One can verify that in the calculations of radiative corrections currently
available~\cite{knrt,cirig01,cuplov} $X_6$ and $K_{12}$ are
always involved through the above combination.
This contribution, related to wave-function renormalisation, 
has the property of being universal, i.e., 
it appears as a multiplicative factor
\be
S_{EW}=1-{1\over2}e^2 (X_6^r -4 K_{12}^r) \,,
\en
in front of the amplitude independently of the specific process considered.
We recover here the universal logarithmically enhanced $\log M_Z$ term
identified by Sirlin~\cite{sirlin82}, but we also get an explicit expression
for the remaining terms.

Let us now consider the problem of perturbative QCD contributions. The
couplings $X_i$ which are related to the difference $VV-AA$, clearly, 
will be essentially unaffected by these. On the contrary, the combination
$X_6^{eff}$ is concerned by such corrections. In fact, if we take into
account the correction proportional to $\alpha_s$ in the asymptotic
behaviour of $\Gamma_{VV}+\Gamma_{AA}$~\cite{sirlin78} 
\be\lbl{alphacor}
(\Gamma_{VV}(k^2)+\Gamma_{AA}(k^2))_{\alpha_s}\sim 
-\,{2\over \pi} {\alpha_s(k^2)\over k^2}\,,
\en
in eq.~\rf{x6int} the integral will diverge. This is to be expected since
the counterterms proportional to $\alpha\,\alpha_s$ have not been implemented.
Inspired by the work of Sirlin~\cite{sirlin82}, one can rather easily
surmount this difficulty. The key point is that the asymptotic behaviour
of $\Gamma_{VV}+\Gamma_{AA}$  is expected to
set in at a scale $\mu_2$ which is much smaller than $M_Z$. As a consequence,
we can rewrite eq.~\rf{x6int}, 
up to very small corrections of order $(\mu_2/M_Z)^2$, 
in terms of an integral in euclidian space with a cutoff at $M_Z$
\bea\lbl{x6intbis}
&&X_6^r(\mu)-4K^r_{12}(\mu)\simeq {1\over32\pi^2}\int_0^{M^2_Z}dx\,[
\Gamma_{VV}(-x)+\Gamma_{AA}(-x)]
\nonumber\\
&&\phantom{X_6^r(\mu)-4K^r_{12}(\mu)\simeq {1\over32\pi^2}\int}
+\facpi \left[ -6\log{M_Z\over\mu} +{5\over2}\right]\ .
\ena
In this form,  it becomes possible to account for the logarithmic terms in
the asymptotic behaviour of $\Gamma_{VV}+\Gamma_{AA}$ without encountering 
any divergence. We will make use of this feature in the next section.

At this point, we have discussed all the LEC's introduced in ref.~\cite{knrt}
except for $X_4$ and $X_7$. Concerning the former, the
corresponding term in the chiral Lagrangian involves only leptons and sources
and bears no relevance for physical low-energy processes. The LEC
$X_7$ has a decomposition given by eq.~\rf{x6x7}. In this expression, the 
LEC $\hat X_7$ parameterises the electromagnetic contribution to the lepton
mass, which is not an observable quantity.

\section{Minimal consistent resonance model}\label{sec:minicons}

\subsection{Estimation of the chiral couplings}

The previous results can be applied to estimate numerically 
the chiral coupling constants which may be of physical relevance. 
One expects that major contributions in the
integrands should come from light, narrow, resonances, which
brings naturally to construct resonance models for the various form-factors.
This idea was put into practice in the case of the 
form-factor $\Pi_{VV-AA}$ a long time ago by Weinberg~\cite{weinVVAA}. He
showed that a minimal model comprising
the $\pi$, $\rho$ and $a_1$ resonances could yield
reasonable results and satisfy the leading QCD asymptotic constraints 
(which determine all the resonance coupling constants in terms of the masses).
In this model $\Pi_{VV-AA}$ reads
\be\lbl{Pivvaa}
\Pi_{VV-AA}(k^2)={\mad \mvd \over k^2 (k^2-\mvd) (k^2-\mad)}\ .
\en
This resonance model was applied to the sum rule calculating
the $\pi^+-\pi^0$ mass difference~\cite{dgmly} and gives a very accurate 
result. The generalisation of this minimal resonance model to the form
factors $F$ and $G$ was discussed in ref.~\cite{bm97}
\be
F(p^2,q^2)={p^2-q^2 + 2(\mad-\mvd)\over 2(p^2-\mvd)(q^2-\mad)}\ ,\quad
G(p^2,q^2)={-q^2 +2\mad\over (p^2-\mvd)(q^2-\mad) q^2}\ ,
\en
and the form factors $\Gamma_{VV}$ and $\Gamma_{AA}$ were discussed in ref.~\cite{knyff01} 
\bea
&&\Gamma_{VV}(k^2,k.r)= {2 k^2-2k.r -c_V\over 2(k^2-\mvd)((r-k)^2-\mvd)} 
\nonumber\ ,\\
&&\Gamma_{AA}(k^2,k.r)= {2 k^2-2k.r -c_A\over 2(k^2-\mad)((r-k)^2-\mad)} 
\ena
(see also refs.~\cite{bijgam,vap} for related work).
The values of $c_V$ and $c_A$ are determined by the Wess-Zumino-Witten
anomalous Lagrangian~\cite{wzw},
\be
c_V= {N_c M_V^4\over  4\pi^2 F_0^2}\ ,\qquad 
c_A= {N_c M_A^4\over 12\pi^2 F_0^2}\ .
\en

Approximating correlators with rational functions is justified in the large-$N_c$ 
limit~\cite{thooft}. But it is not really known whether only retaining the
very first few poles should yield an accurate approximation of the actual
Green functions. One can think of systematically improving
on the minimal model by including more resonance poles together with more
asymptotic constraints (see e.g.~\cite{knyff01}). This interesting possibility
is left for future work, and we stick to the minimal approximation in this paper.

Computing the integrals of sec.~\ref{sec:srules} in the minimal resonance model
is straightforward. If we denote the ratio of the $a_1$ and $\rho$ resonance masses
$z=\mad/\mvd$, we obtain for $X_1$, $\hat X_1$ and $X_3$
\bea\lbl{x1}
&& X_1= -{3\over8}\facpi\left ( \Lz +{c_V z -c_A\over 2\mvd z}\right)
\nonumber\ ,\\
&& \hat X_1= {3\over8}\facpi\left ( -2\LdvZ + \Lz 
  -{c_V z +c_A\over 2\mvd z} +2 \right)
\nonumber\ ,\\
&& X_3^r(\mu)= {3\over2}\facpi\left( \Ldvmu +{\Lz\over z-1} 
-{1\over6}\right)\ .
\ena
For  $X_2^r$ and $\hat X_2^r$ one gets
\bea\lbl{x2}
&& X_2^r(\mu)= {3\over8}\facpi\left( 
\Ldvmu +{\Lz\over z-1} + {5\over 6} \right)\ ,
\nonumber\\
&& \hat X_2^r(\mu)= 
{1\over8}\facpi\left( -10\,\Ldvmuz+7\,\Ldvmu  +{3 (z+1)\Lz\over (z-1)^2}
-{6z\over z-1} +{5\over2}\right)
\nonumber\\
&& \phantom{\hat X_2^r(\mu)= }
-{1\over4} g_{02}^r(\mu_0) + {1\over4} g_{03}^r(\mu_0)\ .
\ena
Finally, the physical combination $X_6-4 K_{12}$ reads
\be
X_6^r(\mu)-4K^r_{12}(\mu)= \facpi\left( 
-8\, \log{M_Z\over M_V} +3\,\Ldvmu + {1\over2}\Lz -{c_V z +c_A\over 4\mvd z} 
+{7\over2}
\right)\ .
\en
This expression accounts for the contribution of the light resonances. 
In the asymptotic region ($k^2> \mu_2^2$ with $\mu_2\simeq 2$ GeV), 
however, our resonance model becomes inaccurate. In particular, while
it reproduces (by construction) the leading asymptotic behaviour 
of $\Gamma_{VV}+\Gamma_{AA}$ 
it does not generate the logarithmic correction proportional to $\alpha_s$
(see eq.\rf{alphacor}.  
We can estimate the modification in the value of $X_6^{eff}$ 
induced by this effect following  ref.~\cite{sirlin78} and 
the discussion in sec.~\ref{sec:secx6}. 
We content ourselves with an unsophisticated leading order
expression for $\alpha_s$. Then, from eqs.~\rf{alphacor}~\rf{x6intbis} 
an analytical evaluation for this correction is obtained
\be
\left( X^{eff}_6\right)_{\alpha_s}\simeq {1\over 4\pi^2\beta_0}\left[
  \log\left(\log{M_Z^2\over\Lambda^2_{QCD}}\right)
- \log\left(\log{\mu_2^2\over\Lambda^2_{QCD}}\right)\right]\ .
\en
In practice, we will use $\beta_0=11-{2\over3}N_f$ with $N_f=4$ and
$\Lambda_{QCD}=206$ MeV which corresponds to $\alpha_s(m^2_\tau)=0.35$
and $\alpha_s(M^2_Z)=0.124$.
The numerical value of $\left(X_6^{eff}\right)_{\alpha_s}$ is shown in table 1.
The table also shows the numerical values of the LEC's generated by the
minimal resonance model (with $z=2$,  $\mu=M_V=0.77\ {\rm GeV}$). 
In the case of $X_6^{eff}$ we observe that
the $\alpha_s$ correction is sizable but the resonance contribution dominates.
Both contributions have the same sign which is opposite to that of the
large logarithm.
\begin{table}[t]
\begin{center}
\begin{tabular}{|c|c|c|c|c|c|c|}\hline
$10^3\,X_1$ & $10^3\,X^r_2$ & $10^3\,X^r_3$ & $10^3\,\tilde X_6^{eff}$&
$10^3(X_6^{eff})_{\alpha_s} $ &
$10^3 \,X_6^{eff} $ \\ \hline
-3.7 & 3.6 & 5.00 & 10.4 & 3.0 & -231.5   \\ \hline 
\end{tabular}
\caption{\sl\small Numerical values of the 
physically relevant LEC's in the minimal
resonance model with $\mu=M_V=0.77$ GeV and $M_A^2/M_V^2=2$. In the case of
$X_6^{eff}$ we show separately the resonance contribution without the large
logarithm (column 4), with the large logarithm (column 6) and in column 5
the perturbative $\alpha_s$ correction (see text).}
\end{center}
\end{table}

\subsection{Examples of applications}

Let us select a few applications of our results for illustrative purposes.
To begin with, let us evaluate the Marciano-Sirlin constant $C_1$ which appears
in the $\pi_{l2}$  decay amplitude~\cite{marciano}. Comparing with the
one-loop amplitude in chiral perturbation theory, Knecht~\emph{et al.}~\cite{knrt}
have derived the decomposition of $C_1$ in terms of LEC's and chiral logarithms
\bea
&&C_1= -4\pi^2\Big[ {8\over3}(K_1^r+K_2^r)+{20\over9}(K_5^r+K_6^r)
\nonumber\\
&&\phantom{C_1= -4\pi^2\Big[}
-{4\over3} X_1 +4(-X_2^r+X_3^r)-(\tilde X^r_6-4K^r_{12})\Big]_{\mu=M^2_\rho}
\nonumber\\
&&\phantom{C_1= -4\pi^2\Big[}
+{Z\over 4}
\left(3 +2\log{M^2_\pi\over M^2_\rho}+\log{M^2_K\over M^2_\rho}\right)
-{1\over2}\ 
\ena
where (following~\cite{cirig01}) 
$\tilde X^r_6$ is defined as $X^r_6$ minus the large logarithm. 
All the LEC's participating in this expression have been estimated on the
basis of the consistent minimal resonance model. The $O(p^2)$ LEC $Z$ is
given in terms of $\langle VV-AA\rangle$ by the Das~\emph{et al.} sum 
rule~\cite{dgmly}, which yields in the minimal resonance model
\be
Z= {3\over2}\facpi {\mvd\over F^2_0}{ z\Lz \over z-1}\simeq 0.92\ .
\en
The sums $K_1+K_2$ and $K_5+K_6$ have been evaluated through the modeling
of a set of QCD 4-point functions~\cite{am}. Combining these results
with the estimates presented in this paper (table 1), we obtain
\be
C_1\simeq -0.93-1.63=-2.56\ 
\en
where the first contribution comes from the LEC's $K_i$ and $X_i$. 
In our opinion, the uncertainty on our estimates of these should not exceed
50\%, which gives for the error on $C_1$
\be
\Delta C_1\simeq 0.5\ .
\en
Our result for $C_1$ lies at the margin of the range 
guessed in ref.~\cite{marciano}, $-2.4\le C_1\le 2.4$. 
It can be applied to extract a slightly more precise value of the pion
decay constant $F_\pi$. Starting from eq.~(21) of ref.~\cite{marciano},
\be
\sqrt2 F_\pi= 130.7 \left( 0.9750\over V_{ud}\right)\pm 0.1 + 0.15\, C_1
\ {\rm MeV}
\en  
and using an updated value for $V_{ud}$ from ref.~\cite{czarneki}, we obtain
\be
F_\pi= 92.2\pm 0.2\ {\rm MeV}\ .
\en

As a second application, let us consider the ratio of the form factors arising
in $K^0_{l3}$ and $K^+_{l3}$ decays. It was noted in ref.~\cite{cirignp04}
that the only unknown input in one-loop chiral perturbation theory
is the LEC $X_1$
\be
{f^{K^+\pi^0}_+(0)\over f^{K^0\pi^-}_+(0)}\Big{\vert}_{ChPT}\equiv
r^{th}_{0+} =1.022 \pm 0.003-16\pi\alpha X_1\ .
\en
Our estimate for $X_1$ induces only very limited changes, 
giving $r^{th}_{0+} =1.023 \pm 0.003$ which remains somewhat incompatible
with the present experimental determination ~\cite{E865,NA48,KTev,KLOE}
$r^{exp}_{0+} =1.038 \pm 0.007$ (see~\cite{neufeldbh} ). Let us stress that the
determination of $X_1$ should be reasonably accurate, since it involves
the difference $VV-AA$ in an integral relation with a rapid convergence.

\section{Conclusions}

In this paper, we have studied the matching of the Standard Model to 
the chiral Lagrangian describing the dynamics of its lightest degrees 
of freedom at low energies. The high-energy dynamics of the Standard
Model is encoded into the low-energy constants (LEC's) which are factors
of local counterterms in the chiral Lagrangian. We have focused on
the LEC's $X_i$ that describe radiative corrections to 
weak semi-leptonic decays.

To determine the connection between these LEC's and the Standard Model,
we have followed a two-step procedure. We started from 
the decay amplitude of a lepton, computed at one loop in the Standard Model,
and we matched it onto the corresponding computation in Fermi theory.
This has allowed us to determine the relevant counterterms 
in the latter theory.  Then comes the second step of our matching procedure, 
from Fermi theory to the chiral effective Lagrangian. 
Thanks to a set of correlators defined within a spurion framework, 
we have related these Fermi counterterms to the chiral LEC's.  
This led us to generate for all the 
$X_i$'s of physical relevance an integral representation 
involving two- and three-point Green functions of 
vector and axial-vector currents defined in chiral QCD.
These can be brought into fairly simple forms involving just
three form-factors: $\Gamma_{VV}$, $\Gamma_{AA}$ and $\Pi_{VV-AA}$. 
Simple but non-trivial relationships among the chiral LEC's are
revealed by these representations.

We dwelt on the case of $X_6$, whose representation involves a combination
of Fermi counterterms left undetermined by the first step of our matching
procedure. This indicated that this LEC should always appear in physical
processes together with another chiral coupling, namely the electromagnetic
LEC $K_{12}$, and we have derived an integral representation for the
physical combination of the two LEC's. In practice this universal term is
dominated by Sirlin's large logarithm. Our approach allows one to identify
the unenhanced terms as well.

Finally, we have estimated the 
values of the $X_i$'s by plugging into the integral
representations a resonance model for the chiral two- and three-point 
Green functions. We have investigated the minimal 
resonance model that satisfies the leading asymptotic QCD constraints, with
poles corresponding to Goldstone boson, vector and axial-vector resonances.
Such a model is expected to yield rather accurate results whenever the 
sum rules are rapidly converging. In the case of the coupling $X_6^{eff}$,
this criterion fails to be satisfied, and we have accounted for the
main correction using perturbative QCD.
Table 1  shows that the resonance  contributions are smaller 
than the large logarithms by approximately a factor of twenty. 

We presented two applications of our results. First, we reexamined the
Marciano-Sirlin constant $C_1$, whose value lies slightly out of the
range guessed in ref.~\cite{marciano}. A second outcome of our analysis
concerns $K_{l 3}$ decays, for which various sets of data exist but are 
barely compatible within experimental errors. A good test of consistency
consisted in the ratio of $K^0_{l3}$ and $K^+_{l3}$ form factors. Our estimate
of $X_1$, based on a resonance model for the $VV-AA$ correlator, is 
too small to account for the discrepancy between experimental data
and chiral perturbation theory.

\medskip
\noindent{\Large\bf Acknowledgments}

We would like to thank Helmut Neufeld and 
Karol Kampf for several judicious remarks.


\begin{thebibliography}{99}

\bibitem{cirignp04}
  V.~Cirigliano, H.~Neufeld and H.~Pichl,
  Eur.\ Phys.\ J.\ C {\bf 35} (2004) 53
  [\texttt{hep-ph/0401173}].

\bibitem{E865}
  A.~Sher {\it et al.},
  Phys.\ Rev.\ Lett.\  {\bf 91} (2003) 261802
  [\texttt{hep-ex/0305042}].

\bibitem{ISTRA}
  O.~P.~Yushchenko {\it et al.},
  Phys.\ Lett.\ B {\bf 589} (2004) 111
  [\texttt{hep-ex/0404030}].
 


\bibitem{NA48}
   A.~Lai {\it et al.} [NA48 Collaboration], 
  Phys.\ Lett.\ B {\bf 602} (2004) 41
  [\texttt{hep-ex/0410059}];
  Phys.\ Lett.\ B {\bf 604} (2004) 1
  [\texttt{hep-ex/0410065}].

\bibitem{KTev}
  T.~Alexopoulos {\it et al.}  [KTeV Collaboration],
  Phys.\ Rev.\ Lett.\  {\bf 93} (2004) 181802
  [\texttt{hep-ex/0406001}];
  Phys.\ Rev.\ D {\bf 70} (2004) 092007
  [\texttt{hep-ex/0406003}].

\bibitem{KLOE}
  P.~Franzini,
  \texttt{hep-ex/0408150}.
 
\bibitem{cirig01}
  V.~Cirigliano, M.~Knecht, H.~Neufeld, H.~Rupertsberger and P.~Talavera,
  Eur.\ Phys.\ J.\ C {\bf 23} (2002) 121
  [\texttt{hep-ph/0110153}].

\bibitem{newrad}
  T.~C.~Andre,
  \texttt{hep-ph/0406006}, 

  V.~Bytev, E.~Kuraev, A.~Baratt and J.~Thompson,
  Eur.\ Phys.\ J.\ C {\bf 27} (2003) 57
  [Erratum-ibid.\ C {\bf 34} (2004) 523]
  [\texttt{hep-ph/0210049}].

\bibitem{kinoshita}
  S.~M.~Berman,
  Phys.\ Rev.\ Lett.\  {\bf 1} (1958) 468,

  T.~Kinoshita,
  Phys.\ Rev.\ Lett.\  {\bf 2} (1959) 477.

\bibitem{sirlin78}
  A.~Sirlin,
  Rev.\ Mod.\ Phys.\  {\bf 50} (1978) 573
  [Erratum-ibid.\  {\bf 50} (1978) 905].

\bibitem{sirlin82}
  A.~Sirlin,
  Nucl.\ Phys.\ B {\bf 196} (1982) 83.

\bibitem{weinberg79}
  S.~Weinberg,
  Physica A {\bf 96} (1979) 327.

\bibitem{gl84}
  J.~Gasser and H.~Leutwyler,
  Annals Phys.\  {\bf 158} (1984) 142.

\bibitem{gl85}
  J.~Gasser and H.~Leutwyler,
  Nucl.\ Phys.\ B {\bf 250} (1985) 465.

\bibitem{dghbook}
  J.~F.~Donoghue, E.~Golowich, B.~H.~Holstein, ``Dynamics of the Standard 
  Model'', Cambridge University Press.

\bibitem{urech95}
  R.~Urech,
  Nucl.\ Phys.\ B {\bf 433} (1995) 234
  [\texttt{hep-ph/9405341}].

\bibitem{knrt}
  M.~Knecht, H.~Neufeld, H.~Rupertsberger and P.~Talavera,
  Eur.\ Phys.\ J.\ C {\bf 12} (2000) 469
  [\texttt{hep-ph/9909284}].

\bibitem{bm97} 
  B.~Moussallam,
  Nucl.\ Phys.\ B {\bf 504} (1997) 381
  [\texttt{hep-ph/9701400}].

\bibitem{dgmly}
  T.~Das, G.~S.~Guralnik, V.~S.~Mathur, F.~E.~Low and J.~E.~Young,
  Phys.\ Rev.\ Lett.\  {\bf 18} (1967) 759.

\bibitem{rusetsky}
  J.~Gasser, A.~Rusetsky and I.~Scimemi,
  Eur.\ Phys.\ J.\ C {\bf 32} (2003) 97
  [\texttt{hep-ph/0305260}].

\bibitem{bl}
  E.~Braaten and C.~S.~Li,
  Phys.\ Rev.\ D {\bf 42} (1990) 3888.

\bibitem{iz}
  C.~Itzykson and J.B.~Zuber,
  ``Quantum field theory'',
  McGraw-Hill (1980), sec. 5-1-3.

\bibitem{egpr}
  G.~Ecker, J.~Gasser, A.~Pich and E.~de Rafael,
  Nucl.\ Phys.\ B {\bf 321} (1989) 311.

\bibitem{savage}
  M.~J.~Savage, M.~E.~Luke and M.~B.~Wise,
  Phys.\ Lett.\ B {\bf 291} (1992) 481
  [\texttt{hep-ph/9207233}].

\bibitem{knyff01}
  M.~Knecht and A.~Nyffeler,
  Eur.\ Phys.\ J.\ C {\bf 21} (2001) 659
  [\texttt{hep-ph/0106034}].

\bibitem{weinbook}
  S.~Weinberg, ``The quantum theory of fields'', 
  Cambridge University Press (1996), vol. 2, sec. 20.5.

\bibitem{dmo}
  T.~Das, V.~S.~Mathur and S.~Okubo,
  Phys.\ Rev.\ Lett.\  {\bf 19} (1967) 859.

\bibitem{cuplov}
  V.~Cuplov and A.~Nehme,
  \texttt{hep-ph/0311274}.

\bibitem{weinVVAA}
  S.~Weinberg,
  Phys.\ Rev.\ Lett.\  {\bf 18} (1967) 507.

\bibitem{bijgam}
  J.~Bijnens, E.~Gamiz, E.~Lipartia and J.~Prades,
  JHEP {\bf 0304} (2003) 055
  [\texttt{hep-ph/0304222}].

\bibitem{vap}
  V.~Cirigliano, G.~Ecker, M.~Eidemuller, A.~Pich and J.~Portoles,
  Phys.\ Lett.\ B {\bf 596} (2004) 96
  [\texttt{hep-ph/0404004}].

\bibitem{wzw}
  J.~Wess and B.~Zumino,
  Phys.\ Lett.\ B {\bf 37} (1971) 95,
  E.~Witten,
  Nucl.\ Phys.\ B {\bf 223} (1983) 422.

\bibitem{thooft}
  G.~'t Hooft,
  Nucl.\ Phys.\ B {\bf 72} (1974) 461.

\bibitem{marciano}
  W.~J.~Marciano and A.~Sirlin,
  Phys.\ Rev.\ Lett.\  {\bf 71} (1993) 3629.

\bibitem{czarneki}
  A.~Czarnecki, W.~J.~Marciano and A.~Sirlin,
  Phys.\ Rev.\ D {\bf 70} (2004) 093006
  [arXiv:hep-ph/0406324].
  

\bibitem{neufeldbh}
  H.~Neufeld,  
  Proceedings of the 337th WE-Heraeus seminar, Bad Honnef, Germany, 
  December 13-17, 2004. ,   
  ``Effective field theories in nuclear, particle and atomic physics''
  J.~Bijnens, U.~G.~Meissner and A.~Wirzba eds, \texttt{hep-ph/0502008}.

\bibitem{am}
  B.~Ananthanarayan and B.~Moussallam,
  JHEP {\bf 0406} (2004) 047
  [\texttt{hep-ph/0405206}].

\end{thebibliography}
\end{document}